\documentclass[1p,sort&compress]{elsarticle}
\usepackage{amsmath,amssymb,amscd,enumerate}
\usepackage{color}
\makeatletter
\def\ps@pprintTitle{%
 \let\@oddhead\@empty
 \let\@evenhead\@empty
 \def\@oddfoot{}%
 \let\@evenfoot\@oddfoot}
\makeatother
\usepackage{hyperref}
\journal{Nuclear Physics B}
\newcommand{\bea}{\begin{eqnarray}}
\newcommand{\eea}{\end{eqnarray}}
\newcommand{\bse}{\begin{subequations}}
\newcommand{\ese}{\end{subequations}}

\newcommand{\noi}{\noindent}

\newcommand{\ba}{\begin{array}}
\newcommand{\ea}{\end{array}}
\newcommand{\balign}{\begin{align}}
\newcommand{\ealign}{\end{align}}

\newcommand{\be}{\beta}
\newcommand{\de}{\delta}
\newcommand{\ep}{\epsilon}
\newcommand{\vep}{\varepsilon}

\newcommand{\la}{\lambda}

\newcommand{\si}{\sigma}

\newcommand{\ze}{\zeta}
\newcommand{\De}{\Delta}

\newcommand{\La}{\Lambda}

\newcommand{\mbf}[1]{\mathbf{#1}}
\newcommand{\bx}{\mathbf x}
\newcommand{\bn}{\mathbf n}

\newcommand{\mbs}[1]{\boldsymbol{#1}}
\newcommand{\bxi}{\mbs\xi}

\newcommand{\mbb}[1]{\mathbb{#1}}

\newcommand{\ms}[1]{\mathsf{#1}}

\newcommand{\mc}[1]{\mathcal{#1}}

\renewcommand{\le}{\leqslant}
\renewcommand{\ge}{\geqslant}
\renewcommand{\leq}{\leqslant}
\renewcommand{\geq}{\geqslant}

\renewcommand{\v}{\vert}

\newcommand{\su}[1]{su($#1$)}

\newcommand\ket[1]{|#1\rangle}

\newcommand{\qbinom}[3]{\genfrac{[}{]}{0pt}{}{\,#1\,}{#2}_{#3}}

\newcommand{\eq}[1]{(\ref{#1})}
\newcommand{\Eq}[1]{Eq.~(\ref{#1})}
\newcommand{\tr}{\operatorname{tr}}

\renewcommand{\mod}{\operatorname{mod}}

\newcommand{\e}{\mathrm{e}}

\newcommand{\RR}{{\mathbb R}}
\newcommand{\Z}{Z^{(m_1,m_2)}}
\newcommand{\cZ}{{\mathcal Z}^{(m_1,m_2)}}
\newcommand{\LaBpm}{\La_{\mathrm B,\pm}^{(m_1,m_2)}}
\newcommand{\LaBspm}{\La_{\mathrm B,\mathrm s,\pm}^{(m_1,m_2)}}

\begin{document}
\begin{frontmatter}
\title{Rational quantum integrable systems of $D_N$ type
with polarized spin reversal operators}
\author[SINP]{B.~Basu-Mallick}
\ead{bireswar.basumallick@saha.ac.in}

\author[SINP]{C.~Datta}
\ead{chitralekha.datta@saha.ac.in}

\author[UCM]{F.~Finkel}
\ead{ffinkel@ucm.es}

\author[UCM]{A.~Gonz\'alez-L\'opez\corref{cor}}
\cortext[cor]{Corresponding author}
\ead{artemio@ucm.es}
  
  \address[SINP]{Theory Group, Saha Institute of Nuclear Physics, 1/AF Bidhan Nagar, Kolkata 700
    064, India}
  \address[UCM]{Departamento de F\'\i sica Te\'orica II, Universidad Complutense de Madrid, 28040
    Madrid, Spain}

  \begin{abstract}
    We study the spin Calogero model of $D_N$ type with polarized spin reversal operators, as well
    as its associated spin chain of Haldane--Shastry type, both in the antiferromagnetic and
    ferromagnetic cases. We compute the spectrum and the partition function of the former model in
    closed form, from which we derive an exact formula for the chain's partition function in terms
    of products of partition functions of Polychronakos--Frahm spin chains of type~$A$. Using a
    recursion relation for the latter partition functions that we derive in the paper, we are able
    to numerically evaluate the partition function, and thus the spectrum, of the $D_N$-type spin
    chain for relatively high values of the number of spins~$N$. We analyze several global
    properties of the chain's spectrum, such as the asymptotic level density, the distribution of
    consecutive spacings of the unfolded spectrum, and the average degeneracy. In particular, our
    results suggest that this chain is invariant under a suitable Yangian group, and that its
    spectrum coincides with that of a Yangian-invariant vertex model with linear energy function
    and dispersion relation.
  \end{abstract}
\end{frontmatter}

\numberwithin{equation}{section}

\section{Introduction}

In a recent paper, a novel type of spin Calogero models and their associated spin chains of
Haldane--Shastry type was introduced~\cite{BBB14}. The distinguishing feature of these models is
that they are constructed using a new representation of the Weyl group of the $BC_N$ root system,
obtained by replacing the standard spin reversal operators by an arbitrarily polarized version
thereof. As shown in the latter reference, these models are exactly solvable for all such
representations and, in particular, the partition function of the spin chains can be exactly
computed using Polychronakos's freezing trick~\cite{Po93,Po94}. In this paper, we shall extend the
above results to spin Calogero models and their corresponding spin chains of Haldane--Shastry type
based on the $D_N$ root system.

In order to present our work in an appropriate context, let us briefly recall the origin and
significance of the latter models. The Haldane--Shastry (HS) spin chain, introduced independently
by these authors in the late eighties~\cite{Ha88,Sh88}, is perhaps the best known example of an
exactly solvable one-dimensional lattice model with long-range interactions. More precisely, this
model describes a circular array of equispaced spins with two-body interactions inversely
proportional to the square of the (chord) distance between the spins. The motivation for
introducing this chain was the construction of a simple model with an exact ground state given by
the $U\to\infty$ limit of Gutzwiller's variational wavefunction for the ground state of the
one-dimensional Hubbard model~\cite{Gu63,GV87,GJR87}. Over the years, the HS chain has appeared in
many areas of interest both in Physics and Mathematics, such as fractional statistics and
one-dimensional anyons~\cite{Ha91b,GS05,Gr09,HHTBP92}, quantum entanglement~\cite{GSFPA10},
characterization of integrability vs.~quantum chaos~\cite{FG05,BB06,BFGR08epl,BFGR09power},
quantum integrability via the asymptotic Bethe ansatz~\cite{Ha91,Ka92,HH93}, Yangian quantum
groups~\cite{BGHP93,HHTBP92,BS96,KKN97}, and conformal field theory~\cite{Ha91,BBS08,CS10,NCS11}.

One of the key properties of the HS chain ---already noted by Haldane and Shastry in their
original papers--- is its intimate connection with the scalar (trigonometric) Sutherland
model~\cite{Su71,Su72}. This connection was subsequently elucidated by Polychronakos in
Ref.~\cite{Po93}, who showed how to derive the HS chain from the spin Sutherland
model~\cite{HH92,HW93,MP93} by a technique that he called the ``freezing trick''. The main idea
behind this technique is to note that when the coupling constant in the spin Sutherland model goes
to infinity the particles tend to concentrate on the coordinates of the (essentially unique)
minimum of the scalar part of the interaction potential, which are precisely the sites of the HS
chain. Thus, in this limit the dynamical and the spin degrees of freedom decouple, and the latter
are governed by the chain's Hamiltonian. Using this idea it is straightforward, for instance, to
obtain the first integrals of the HS chain from their well-known counterparts for the spin
Sutherland model. In fact, Polychronakos showed that applying the same procedure to the (rational)
Calogero model~\cite{Ca71} and its spin version~\cite{MP93} one obtains an integrable chain with
non-equispaced sites and long-range interactions inversely proportional to the distance between
the spins~\cite{Po93}. The spectrum of this chain ---known in the literature as the
Polychronakos--Frahm (PF) chain--- was first studied numerically by Frahm~\cite{Fr93} and then
exactly computed by Polychronakos~\cite{Po94}, who derived an exact formula for the partition
function by means of the freezing trick. On the other hand, the partition function of the HS chain
was only evaluated more than a decade later by some of the authors~\cite{FG05}.

Both the Sutherland and Calogero models (and their corresponding HS and PF spin chains) are
associated with the $A_{N-1}$ root system, where $N$ is the number of particles. Indeed, in these
models the interactions only depend on the difference between the coordinates, and the spin
operators appearing in the Hamiltonian are permutation operators, and thus generate a realization
of the Weyl group of $A_{N-1}$ type. In fact, there are versions of the Sutherland and Calogero
models associated to any (extended) classical root system~\cite{OP83}. Among these systems, those
of $BC_N$, $B_N$, $C_N$ and $D_N$ type are by far the most studied in the literature, since they
make it possible to construct integrable models with an arbitrary number of particles. By applying
the freezing trick to the spin version of these models one obtains the corresponding
generalizations of the HS and PF chains, that we shall collectively refer to as spin chains of HS
type~\cite{BPS95,YT96,EFGR05,BFGR08,BFG09,BFG11,BFG13}. One of the fundamental features of the
$BC_N$ root system and its $B_N$ and $C_N$ degenerations is the fact that its Weyl algebra
contains a family of reflection operators $S_i$ ($i=1,\dots,N$) satisfying $S_i^2=1$. (In the case
of the $D_N$ root system, the Weyl group only contains products $S_iS_j$ with $i\ne j$.) In the
spin chains studied in Refs.~\cite{BPS95,YT96,EFGR05,BFGR08,BFG09,BFG11,BFG13}, the operators
$S_i$ are represented by spin reversal operators $P_i$ (acting on the Hilbert space of the $i$-th
particle), but this is by no means the only possible choice. As a matter of fact, in the novel
version of the spin Calogero model of $BC_N$ type and its corresponding (PF) chain introduced in
Ref.~\cite{BBB14}, the operators $S_i$ are represented instead by arbitrarily polarized spin
reversal operators (PSRO) $P_i^{(m_1,m_2)}$, which act as the identity on the first $m_1$ elements
of the spin basis and as minus the identity on the rest. These operators are equivalent under a
similarity transformation to the usual spin reversal operators $P_i$ only when $m_1=m_2$ or
$m_1=m_2\pm1$, i.e., when there is minimal polarization. For the remaining values of the discrete
parameters $m_1$ and $m_2$, the systems constructed in the latter reference differ from the
standard Calogero and PF models of $BC_N$-type. In particular, when $m_1$ or $m_2$ are zero, the
corresponding spin dynamical model reduces to the $\mathrm{su}(m)$-invariant extension of the
Calogero model studied by Simons and Altshuler~\cite{SA94}; see also~\cite{FGGRZ01}.

In this paper we introduce the spin Calogero model of $D_N$-type with PSRO and its corresponding
spin chain of HS type, i.e., the PF chain of $D_N$ type with PSRO. As explained in
Ref.~\cite{BFG09}, these models are singular limits of their corresponding $BC_N$ counterparts, so
that their spectrum cannot be obtained by setting to zero the parameter $\be$ in the latter models
(cf.~Eqs.~\eqref{ham} and~\eqref{l14}). This is also apparent at the level of the Hilbert space,
which is the direct sum of the Hilbert spaces of two $BC_N$ models with opposite chiralities.
Thus, the models studied in this paper are not limiting cases of their $BC_N$ versions in
Ref.~\cite{BBB14}.

Our main result is the derivation of a closed-form expression for the partition function of the PF
chain of $D_N$ type with PSRO in terms of products of partition functions of type-$A$ PF chains.
Our approach is based on the computation of the spectrum and partition function of the
corresponding spin Calogero model, from which the chain's partition function follows by a standard
freezing trick argument. The structure of this partition function turns out to be more involved
than that of its $BC_N$ counterpart. In particular, it is not manifest that it is a polynomial in
$q\equiv\e^{-1/(k_{\mathrm B}T)}$, as follows from the freezing trick. Using the explicit
expression for the partition function, we shall study several global properties of the chain's
spectrum, such as the behavior of the level density and the average degeneracy when the number of
spins tends to infinity. In particular, the fact that the number of distinct levels grows
polynomially with the number of spins suggests that this model is isospectral to a
Yangian-invariant vertex model of the kind studied in Ref.~\cite{BBH10}.

The paper is organized as follows. In Section~\ref{sec.consm} we recall the definition and main
properties of the polarized spin reversal operators $P_i^{(m_1,m_2)}$, and construct the
Hamiltonians of the $D_N$-type spin Calogero model with PSRO and its associated spin chain.
Section~\ref{sec.specPF} is devoted to the derivation of the closed-form expression of the chain's
partition function, as explained in the previous paragraph. Using this expression, in
Section~\ref{sec.stat} we analyze several global properties of the spectrum, providing strong
numerical evidence of the Gaussian character of the level density when the number of spins is
large enough. In Section~\ref{sec.ferro} we extend the above results to the ferromagnetic version
of the models under consideration. The paper ends with a brief section summarizing our main
results and presenting our conclusions, and a short technical Appendix establishing a useful
recursion relation for the partition function of the PF chain of type $A_{N-1}$.

\section{Construction of the models}\label{sec.consm}

For the purpose of describing the $D_N$-type Calogero model with polarized spin reversal operators,
it is convenient to briefly summarize the construction of its $BC_N$ counterpart \cite{BBB14}. To
this end, let
\begin{equation} \mc{S}= \big{\langle} \ket{s_1,\ldots,s_N} ~ \big{\vert} s_i \in \{ 1,2,
\ldots , m \} \big{\rangle} \, .
\label{l2}
\end{equation}
denote the internal spin space for $N$ particles. As usual, the action of the spin exchange
operator $P_{ij}$ on $\mc{S}$ is defined as
\begin{equation}
P_{ij}\ket{s_1, \ldots ,s_i, \ldots ,s_j, \ldots ,s_N}
=\ket{s_1, \ldots ,s_j, \ldots ,s_i, \ldots ,s_N} \, .
\label{l3}
\end{equation} Let us denote the PSRO associated with the $i$-th particle as $P_{i}^{(m_1,m_2)}$, where
$m_1$ and $m_2$ are two nonnegative integers satisfying the relation $m_1+m_2=m$. The
action of $P_{i}^{(m_1,m_2)}$ on $\mc{S}$ is given by~\cite{BBB14}
\begin{equation}
P_i^{(m_1,m_2)}\ket{s_1, \ldots ,s_i, \ldots ,s_N}
=(-1)^{f(s_i)}\ket{s_1, \ldots ,s_i, \ldots ,s_N} \, ,
\label{l4}
\end{equation}
where 
\begin{equation}
 f(s_i) = \begin{cases}
0,  & 1\le s_i\le m_1\\
1, &  m_1+1\le s_i\le m_1+m_2.
\end{cases}
\label{l5}
\end{equation}
In terms of these operators, the Hamiltonian of the $BC_N$-type Calogero model with PSRO is
defined as~\cite{BBB14}
\begin{equation}
H_{\mathrm B,\ep}^{(m_1,m_2)}=-\sum_i
\frac{\partial^2}{\partial x_i^2}+a\sum\limits_{i\neq j}
\left( \frac{a+P_{ij}}{(x_{ij}^-)^2}+\frac{a+\widetilde
{P}_{ij}^{(m_1,m_2)}}{(x_{ij}^+)^2}\right)
+\beta a \sum_{i}\frac{\beta a-\ep P_i^{(m_1,m_2)}}{x_i^2}
+\frac{a^2}{4}r^2 \, , 
\label{ham}
\end{equation}
where the sums run from $1$ to $N$, $a> \frac{1}{2}$, $\beta > 0$,
$\ep=\pm 1$, 
$x_{ij}^{\pm}=x_i\pm x_j$, $r^2=\sum_i x_i^2$, and
\[
\widetilde{P}_{ij}^{(m_1,m_2)}=P_i^{(m_1,m_2)}P_j^{(m_1,m_2)}P_{ij}.
\]
It can be shown that when $m$ is even (resp.~odd) and $m_1=m_2$ (resp.~$m_1=m_2\pm1$), the PSRO in
\eq{l4} is equivalent via a similarity transformation to the usual spin reversal operator $P_i$,
which changes $s_i$ into $m-s_i+1$. As a result, for these special choices of $m_1$ and $m_2$, the
Hamiltonian \eq{ham} reduces to that of the standard $\mathrm su(m)$ spin Calogero model of $BC_N$
type studied in Ref.~\cite{BFGR08}. As mentioned in the Introduction, another interesting special
case is $m_1=m$, $m_2=0$, for which the Hamiltonian \eq{ham} reduces to the Simons--Altshuler
extension of the Calogero model.
 
Since $H_{\mathrm B,\ep}^{(m_1,m_2)}$ contains the discrete parameters $m_1$, $m_2$ and $\ep$, it
is natural to inquire whether there exists any relation between models~\eq{ham} with different
sets of parameters. In fact, we shall now show that $H_{\mathrm B,\ep}^{(m_1,m_2)}$ is equivalent
to $H_{\mathrm B,-\ep}^{(m_2,m_1)}$ through a unitary transformation. To this end, consider the
unitary operator $T$ whose action on the spin space $\mc{S}$ is given by
\begin{equation}
T\ket{s_1,\ldots,s_i,\ldots,s_N}=\ket{s_1^{\prime},\ldots,s_i^{\prime},\ldots, s_N^{\prime}},
\label{l6}
\end{equation}
where
\begin{equation}
 s_i^{\prime} =
\begin{cases}
s_i+m_1, & 1\le s_i\le m_2\\
s_i-m_2, & m_2+1\le s_i\le m_1+m_2.
\end{cases}
\label{l7}
\end{equation}
Using Eqs.~\eq{l4} and \eq{l6} we easily obtain
\begin{equation}
{T^\dagger} P_i^{(m_1,m_2)}T\ket{s_1, \ldots ,s_i, \ldots ,s_N}
=(-1)^{f(s_i^{\prime})}\ket{s_1, \ldots ,s_i, \ldots ,s_N} \, . 
\label{l8}
\end{equation}
From Eq.~\eq{l7} it follows that $s_i^{\prime}\in \{m_1+1, \ldots,m_1+m_2\}$
for $1\le s_i\le m_2$ and 
$s_i^{\prime}\in \{1, \ldots , m_1\}$ for 
$m_2+1\le s_i\le m_2+m_1$, so that
\begin{equation}
 f(s_i^{\prime}) =
\begin{cases}
1, & 1\le s_i\le m_2\\
0, & m_2+1\le s_i\le m_2+m_1.
\end{cases}
\label{l9}
\end{equation}
Equations \eq{l8} and \eq{l9} clearly imply that
\begin{equation} T^\dagger P_i^{(m_1,m_2)}T=-P_i^{(m_2,m_1)} \, .
\label{l10}
\end{equation}
It is also obvious from Eqs. \eq{l3} and \eq{l6} that
\begin{equation}
T^\dagger P_{ij}T=P_{ij} \, .
\label{l11}
\end{equation}
From Eqs.~\eq{ham}, \eq{l10} and \eq{l11} we readily obtain
\begin{equation}
T^\dagger H_{\mathrm B,\ep}^{(m_1,m_2)}T=H_{\mathrm B,-\ep}^{(m_2,m_1)},
\label{l12}
\end{equation} as claimed. In view of the above relation, it suffices to study the Hamiltonian~\eq{ham} in
the case $\ep=1$. However, in the paper we shall intentionally keep the parameter $\ep$
in $H_{\mathrm B,\ep}^{(m_1,m_2)}$ in order to facilitate the comparison with its $D_N$
counterpart that we shall introduce below.

Due to the nature of the singularities of the Hamiltonian $H_{\mathrm B,\ep}^{(m_1,m_2)}$,
its configuration space can be taken as one of the Weyl chambers of the $BC_N$ root system, i.e.,
one of the maximal open subsets of $\RR^N$ on which the functions $x_i\pm x_j$ and $x_i$ have
constants signs. We shall choose this configuration space as the principal Weyl chamber
\begin{equation}
{C}^{(\mathrm B)}=\left\{\bx\in\RR^N: 0< x_1
  < x_2< \cdots <x_N \right\},
\label{l13}
\end{equation}
where $\bx\equiv(x_1, \ldots , x_N)$. The Hamiltonian $H_{\mathrm B,\ep}^{(m_1,m_2)}$ is thus
defined on an appropriate dense subset of the Hilbert space $L^2(C^{(B)})\otimes\mc S$. When
$\ep=1$, the spectrum of $H_{\mathrm B,\ep}^{(m_1,m_2)}$ was computed in Ref.~\cite{BBB14} by
constructing a suitable (non-orthogonal) basis of this Hilbert space in which this
Hamiltonian acts triangularly.

As explained in the latter reference, from the spin dynamical model~\eqref{ham} one can construct
a PF chain of $BC_N$ type with PSRO by applying the freezing trick. The Hamiltonian of this chain
is given by
\begin{equation}  
\mc{H}_{\mathrm B,\ep}^{(m_1,m_2)}=\sum\limits_{i\neq j}
\left[ \frac{1+P_{ij}}{(\zeta_i-\zeta_j)^2}+\frac{1+\widetilde
{P}_{ij}^{(m_1,m_2)}}{(\zeta_i+\zeta_j)^2}\right]+
\beta \sum_i
\frac{1 -\ep P_i^{(m_1,m_2)}}{\zeta_i^2} \, , 
\label{l14}
\end{equation} where the lattice sites $\zeta_i$ are related to the zeros $y_i$ of the Laguerre polynomial
$L_N^{\beta-1}$ by $y_i=\zeta_i^2/2$. The exact partition function of the chain~\eqref{l14} has also been
computed in Ref.~\cite{BBB14} by exploiting its connection with the spin dynamical model~\eq{ham}.

The Hamiltonian $H^{(m_1,m_2)}$ of the $D_N$-type spin Calogero model with PSRO is naturally
defined by dropping the term related to the roots $x_i$ in~$H_{\mathrm B,\ep}^{(m_1,m_2)}$,
i.e., by setting $\beta=0$ in~\Eq{ham}. We thus obtain
\begin{equation} H^{(m_1,m_2)}=-\sum_i
\frac{\partial^2}{\partial x_i^2}+a\sum\limits_{i\neq j}
\left[\frac{a+P_{ij}}{(x_{ij}^-)^2}+\frac{a+\widetilde {P}_{ij}^{(m_1,m_2)}}{(x_{ij}^+)^2}\right]
+\frac{a^2}{4}r^2 \, .
\label{l15}
\end{equation}
It should be noted that, unlike its $BC_N$ counterpart, the latter Hamiltonian does not depend on
$\ep$. Just as before, from Eqs.~\eq{l10} and~\eq{l11} it follows that $H^{(m_2,m_1)}$ is
unitarily equivalent to $H^{(m_1,m_2)}$ under $T$:
\begin{equation}
T^\dagger {H}^{(m_1,m_2)}T=H^{(m_2,m_1)} \,. 
\label{l16}
\end{equation}
Thus we can impose without loss of generality the restriction $m_1\geq m_2$. Consequently, for any
given $m$ one can construct $\lfloor m/2+1\rfloor$ inequivalent spin Calogero models of $D_N$ type
with PSRO, where $\lfloor\,\cdot\,\rfloor$ denotes the integer part. Among these models, only
those with $m_1=m_2$ (for even $m$) or $m_1=m_2+1$ (for odd $m)$ are unitarily equivalent to the
$\mathrm{su}(m)$ spin Calogero model of $D_N$ type with standard spin reversal operators
introduced in Ref.~\cite{BFG09}.

As is the case with the latter model, the configuration space $C$ of the Hamiltonian~\eq{l15} can
be taken as one of the maximal open subsets of $\RR^N$ on which the linear functionals $x_i\pm x_j$
have constant signs. We shall again take $C$ as the principal Weyl chamber of the $D_N$ root
system, namely
\begin{equation}      
  C=\left\{\bx\in\RR^N:
    \v x_1 \v < x_2< \cdots < x_N\right\}.
\label{l17}
\end{equation}
Note that this configuration space contains its $BC_N$ counterpart \eq{l13} as a subset. As
before, the Hamiltonian~\eq{l15} is defined on a suitable dense subspace of the Hilbert
space $L^2(C)\otimes \mc{S}$.

\begin{figure}[h]
      \centering
      \includegraphics[width=8cm]{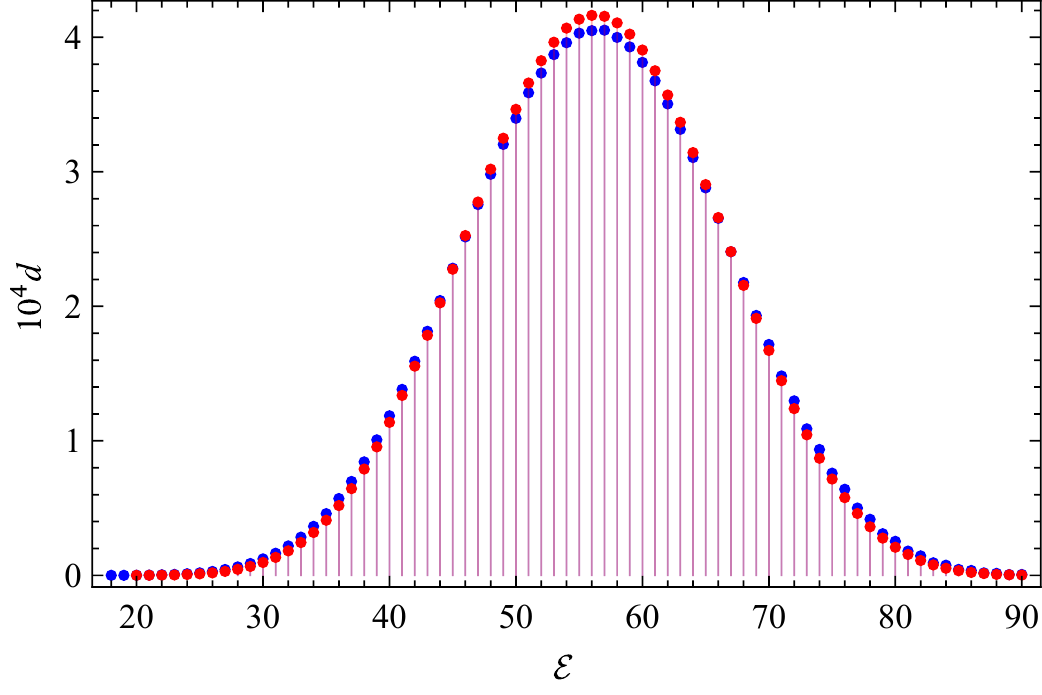}
      \caption{Degeneracy $d$ (in units of $10^4$) versus energy $\mathcal E$ of the $D_N$
        chain~\eqref{l21} with $m_1=3$, $m_2=1$ (blue), compared to the $\mathrm{su}(4)$
        $D_N$-type PF chain with standard time-reversal operators (red), for $N=10$ spins.}
      \label{fig.31-22}
    \end{figure}

We shall next explain in detail how to construct the $D_N$-type PF chain with PSRO associated to
the spin dynamical model~\eqref{l15} by means of Polychronakos's freezing trick. To begin with,
note that the Hamiltonian $H^{(m_1,m_2)}$ can be decomposed as
\begin{equation}\label{HHsccH}
H^{(m_1,m_2)}= H^{\mathrm{sc}}+a\hat{\mc{H}}^{(m_1,m_2)}(\bx)\,,
\end{equation}
where
\begin{equation}
H^{\mathrm{sc}}=-\sum_i\frac{\partial^2}
{\partial x_i^2}+a(a-1)\sum\limits_{i\neq j}
\left[\frac{1}{(x_{ij}^-)^2}+\frac{1}{(x_{ij}^+)^2}\right] 
+\frac{a^2}{4}r^2
\label{l20}
\end{equation}
is the Hamiltonian of the scalar $D_N$ Calogero model and
\begin{equation}\label{hatcH}
\hat{\mc{H}}^{(m_1,m_2)}(\bx)=\sum_{i\neq j} \left[\frac{1+P_{ij}}{(x_{ij}^-)^2}+\frac{1+\widetilde
      {P}_{ij}^{(m_1,m_2)}}{(x_{ij}^+)^2}\right]
\end{equation}
is a spin-dependent multiplication operator. On the other hand, in the strong coupling limit
$a\rightarrow \infty$ the coefficient of the dominant term (of order $a^2$) in the Hamiltonian
\eq{l15} is given by
\begin{equation}
U(\mbf{x})=\sum\limits_{i\neq j}
\left[\frac{1}{(x_{ij}^-)^2}+\frac{1}{(x_{ij}^+)^2}\right]
+\frac{r^2}{4} \,.
\label{l18}
\end{equation}
Hence as $a\to\infty$ the particles concentrate at the coordinates $\xi_i$ of the unique minimum
$\mbs{\xi}$ of the potential $U(\mbf{x})$ in the configuration space $C$~\cite{CS02}, and the
coordinate degrees of freedom of $H^{(m_1,m_2)}$ decouple from the internal ones. By \Eq{HHsccH},
in this limit the eigenvalues of $H^{(m_1,m_2)}$ are approximately given by
\begin{equation}
E_{ij}\simeq E_i^{\mathrm{sc}}+a\mc{E}_j,
\label{m1}
\end{equation}
where $E_i^{\mathrm{sc}}$ and $\mc{E}_j$ are two arbitrary eigenvalues of $H^{\mathrm{sc}}$ and
\begin{equation}
  \mc{H}^{(m_1,m_2)}\equiv \hat{\mc{H}}^{(m_1,m_2)}(\bxi)=\sum_{i\neq j}
  \left[\frac{1+P_{ij}}{(\xi_i-\xi_j)^2}+\frac{1+\widetilde
      {P}_{ij}^{(m_1,m_2)}}{(\xi_i+\xi_j)^2}\right].
\label{l21}
\end{equation}
We shall take Eq.~\eqref{l21} as the precise definition of the Hamiltonian of the $D_N$-type PF
chain with PSRO. In fact, using Eqs.~\eq{l10} and \eq{l11}, it is easy to show that the Hamiltonians
$\mc{H}^{(m_1,m_2)}$ and $\mc{H}^{(m_2,m_1)}$ are related by
\begin{equation}
T^\dagger \mc{H}^{(m_1,m_2)}T=\mc{H}^{(m_2,m_1)} \, . 
\label{l21a}
\end{equation}
Thus, we may assume without loss of generality that $m_1\geq m_2$, so that there are again
$\lfloor m/2+1\rfloor$ inequivalent PF chains of $D_N$ type with PSRO. Since the sites of these
chains depend only on the scalar potential~\eqref{l18}, the above models reduce to the
$\mathrm{su}(m)$ PF chain of $D_N$ type with standard spin reversal operators~\cite{BFG09} when
$m_1=m_2$ (for even $m$) or $m_1=m_2+1$ (for odd $m$). See, e.g., Fig.~\ref{fig.31-22} for a
comparison of the spectra of the $D_N$ chain with PSRO~\eqref{l21} with $m_1=3,m_2=1$ and the
$\mathrm{su}(4)$ $D_N$-type PF chain with standard time-reversal operators (corresponding to
$m_1=m_2=2$) for $N=10$ spins.

\begin{figure}[h]
      \centering
      \includegraphics[width=8cm]{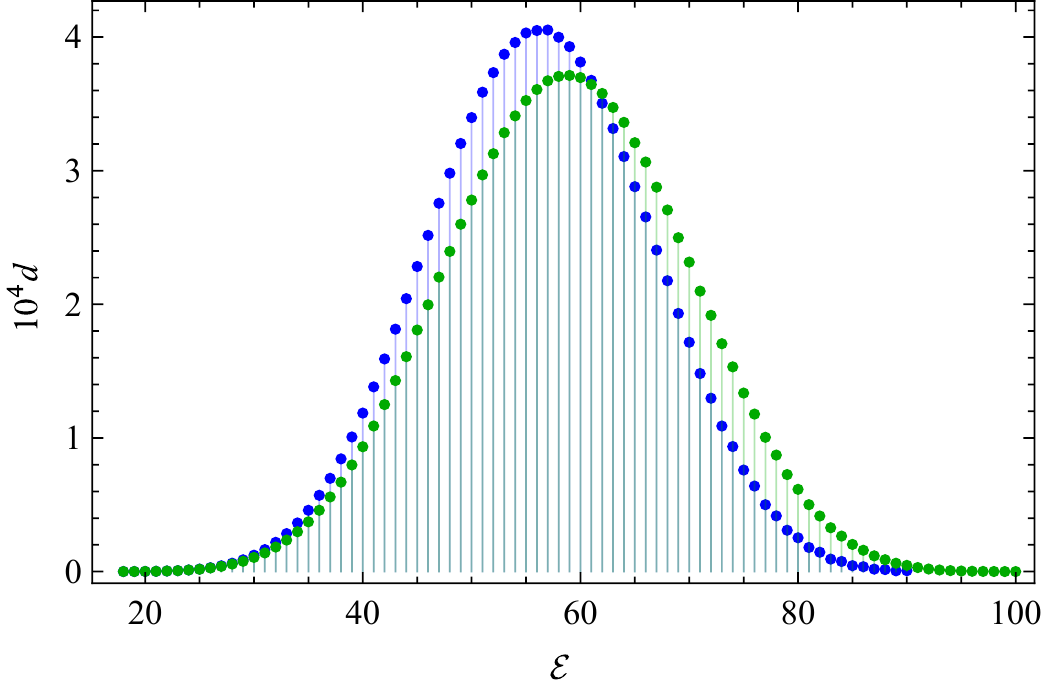}
      \caption{Degeneracy $d$ (in units of $10^4$) versus energy $\mathcal E$ of the $D_N$
        chain~\eqref{l21} with $m_1=3$, $m_2=1$ and $N=10$ spins (blue), compared to its $BC_N$
        counterpart in Eq.~\eqref{l14} (green; recall that the spectrum of the latter chain does
        not depend on $\be$)}
      \label{fig.bd}
    \end{figure}

    A brief remark on the relation between the $D_N$ and $BC_N$ spin chains with PSRO in
    Eqs.~\eqref{l21} and~\eqref{l14} is now in order. As shown in~\cite{BFG09}, the lattice sites
    of the former chain are given by $\xi_1=0$ and $\xi_i=\sqrt{2y_{i-1}}$ ($2\le i\le N$), where
    $y_k>0$ denotes the $k$-th root of the generalized Laguerre polynomial $L_{N-1}^1$. From the
    well-known identity $N L_N^{-1}(y)=-yL_{N-1}^1(y)$ and the previous characterization of the
    sites $\ze_i$ of the $BC_N$ chain~\eqref{l14}, it immediately follows that
    $\mbs{\xi}=\lim_{\beta\to 0}\mbs{\zeta}$. Although one may naively think that the Hamiltonian
    $\mc{H}^{(m_1,m_2)}$ is simply the $\be\to0$ limit of its $BC_N$ counterpart
    $\mc{H}_{\mathrm B,\ep}^{(m_1,m_2)}$, this is certainly not the case. The point is that,
    although the roots $\ze_i$ with $2\le i\le N$ tend to finite nonzero limits when $\be\to0$,
    the first root $\ze_1$ tends to $0$ in this limit. As a consequence, the $i=1$ term of the
    last sum in Eq.~\eqref{l14} need not vanish as $\be\to0$, and in fact it can be
    shown~\cite{BFG09} that
\begin{equation}
\lim_{\beta \rightarrow 0} \frac{\beta}{\ze_1^2}=\frac N2. 
\label{l24}
\end{equation}
Letting $\be\to0$ in Eq.~\eqref{l14} and using the latter identity we immediately obtain
\begin{equation}
\lim_{\beta \rightarrow 0}\mc{H}_{\mathrm B,\ep}^{(m_1,m_2)} 
=\mc{H}^{(m_1,m_2)}+\frac{N}{2}\big(1-\ep P_1^{(m_1,m_2)}\big)\,. 
\label{l25}
\end{equation}
Thus, the $\beta\to0$ limit of the Hamiltonian $\mc{H}_{\mathrm B,\ep}^{(m_1,m_2)}$ differs from
its $D_N$ counterpart $\mc{H}^{(m_1,m_2)}$ by the surface term or impurity interaction
$N\big(1-\ep P_1^{(m_1,m_2)}\big)/2$. It is easy to see that this term vanishes only for $\ep=1,$
$m_1=m$, $m_2=0$ (or, equivalently, $\ep=-1$, $m_1=0$, $m_2=m$). For any other choice of $m_1$ and
$m_2$, this surface term is nonzero and does not commute with the Hamiltonian
$\mc{H}^{(m_1,m_2)}$. Thus, except in the previously noted special cases, the spectrum of
$\mc{H}^{(m_1,m_2)}$ cannot be obtained from that of $\mc{H}_{\mathrm B,\ep}^{(m_1,m_2)}$ by
taking the $\be\to0$ limit. This fact is illustrated in Fig.~\ref{fig.bd}, which shows that the
spectra of these chains with $m_1=3$, $m_2=1$ and $N=10$ spins are clearly different.

\section{Spectrum and partition function}\label{sec.specPF}

In this section, we shall compute in closed form the spectrum and partition function of the spin
Calogero model of $D_N$ type with PSRO in Eq.~\eqref{l15}. This will enable us to compute the
partition function $\cZ$ of the $D_N$-type PF chain with PSRO~\eqref{l21} by a standard freezing
trick argument. Indeed, from Eq.~\eqref{m1} it is straightforward to derive the following exact
formula for $\cZ$ in terms of the partition functions $\Z$ and $Z$ of the spin dynamical
model~\eqref{l15} and of its scalar counterpart~\eqref{l20}:
\begin{equation}
\cZ(T)
=\lim_{a\to\infty}\frac{\Z(aT)}{Z(aT)}.
\label{m2}
\end{equation}
Since $Z$ has already been computed in Ref.~\cite{BFG09}, Eq.~\eqref{m2} provides an
effective way of evaluating $\cZ$ once $\Z$ is known.

The key idea for deriving the spectrum of the spin Hamiltonian~\eqref{l15} is to observe that it
can be obtained by applying a suitable projection to a simpler differential-difference operator
$H'$ acting on scalar functions. The spectrum of $H'$ can be readily computed by constructing a
(non-orthogonal) basis of its Hilbert space on which this operator acts triangularly. The spectrum
of $H^{(m_1,m_2)}$ is then easily determined by projecting onto the Hilbert space of the latter
operator.

More precisely, the auxiliary operator $H'$ is given by~\cite{BFG09}
\begin{equation}
  H^{\prime}=-\sum_i
  \frac{\partial^2}{\partial x_i^2}+a\sum_{i\ne j}
  \left[\frac{a}{(x_{ij}^-)^2}(a-K_{ij})
    +\frac{a}{(x_{ij}^+)^2}(a-\tilde{K}_{ij}) \right]
  +\frac{a^2}{4}r^2 \, ,
\label{m3}
\end{equation}
where $K_{ij}$ and $K_i$ are coordinate exchange and sign reversing operators, 
defined by 
\begin{eqnarray}
&&K_{ij}f(x_1, \ldots ,x_i, \ldots ,x_j, \ldots ,x_N)
=f(x_1, \ldots ,x_j, \ldots ,x_i, \ldots ,x_N) \, ,\nonumber\\
&&K_if(x_1, \ldots ,x_i, \ldots , x_N)=f(x_1, \ldots , -x_i, \ldots ,
x_N) \, , \nonumber
\label{m4}
\end{eqnarray}
and $\widetilde{K}_{ij}\equiv K_iK_jK_{ij}$. The domain of the operator $H'$ is of course a
suitable dense subset of the Hilbert space $L^2(\RR^N)$. The operator $H'$ can be expressed in
terms of the $D_N$-type rational Dunkl operators~\cite{Du98}
\begin{equation}\label{m6}
J_i^-= \frac{\partial}{\partial x_i}+a\sum_{j;j\ne i}
\left[\frac{1}{x_{ij}^-}(1-K_{ij})+
\frac{1}{x_{ij}^+}(1-\tilde{K}_{ij})\right]
\end{equation}
as~\cite{FGGRZ01b}
\begin{equation}
H'=\rho(\mbf{x})
\left[-\sum\limits_{i}(J_i^-)^2+
a\sum\limits_{i}x_i\frac{\partial}{\partial x_i}+E_0 \right] 
\rho(\mbf{x})^{-1}\,,
\label{m5}
\end{equation}
where
\[
\rho(\mbf{x})=\e^{-\frac{a}{4}r^2}\prod\limits_{i< j}\v x_i^2-x_j^2 \v^a
\label{l22}
\]
is the ground state of the scalar Calogero model of $D_N$-type and
\begin{equation}\label{m7}
E_0=Na\big(a(N-1)+\tfrac{1}{2}\big)
\end{equation}
is its ground-state energy. A basis of this Hilbert space on which $H'$ acts triangularly is
provided by the functions
\begin{equation}
\phi_{\mbf{n}}(\mbf {x})
=\rho(\mbf{x})\prod\limits _i x_i^{n_i} \, , \qquad \mbf{n}\equiv(n_1,\dots,n_N),
\label{m8}
\end{equation}
where the $n_i$'s are arbitrary non-negative integers. Indeed, since $J_i^-$ lowers the degree
$|\mathbf n|\equiv n_1+\dots+n_N$ of any monomial $\prod_ix_i^{n_i}$, from Eqs.~\eqref{m6}
and~\eqref{m5} it immediately follows that
\begin{equation}
H'\, \phi_{\mbf{n}}(\mbf{x})
= E_{\mbf{n}}' \, \phi_{\mbf{n}}(\mbf{x})
+\sum\limits _{\v {\mbf m} \v< \v \mbf{n} \v}c_{\mbf{m}\mbf{n}} 
\, \phi_{\mbf m}(\mbf{x}) \, ,
\label{m9}
\end{equation}
where the coefficients $ c_{\mbf{m}\mbf{n}}$ are real constants and
\begin{equation}
E_{\mbf{n}}'= a|\mbf{n}| + E_0 \,.
\label{m10}
\end{equation}
As the diagonal elements of any upper triangular operator coincide with its eigenvalues, the
spectrum of $H'$ is given by \Eq{m10}.

The spectrum of the spin Hamiltonian $H^{(m_1,m_2)}$ can be derived from that of $H'$ by noting
that these Hamiltonians are formally related by
\begin{equation}
H^{(m_1,m_2)}=H^{\prime}|_{K_{ij}\rightarrow-P_{ij}, \, K_iK_j\rightarrow 
P_i^{(m_1,m_2)}P_j^{(m_1,m_2)}}. 
\label{m15}
\end{equation}
In order to take advantage of this observation, we introduce the operator $\Lambda^{(m_1,m_2)}$
projecting the Hilbert space $L^2(\RR^N)\otimes\mathcal S$ onto states that are antisymmetric
under particle permutations and symmetric under the action of
$K_iK_jP_i^{(m_1,m_2)}P_j^{(m_1,m_2)}$ for any $i\ne j$. In other words, the projector
$\Lambda^{(m_1,m_2)}$ is determined by
\begin{equation}\label{piLaD}
  \pi_{ij}\Lambda^{(m_1,m_2)}=-\Lambda^{(m_1,m_2)},\qquad
  \pi_i^{(m_1,m_2)}\pi_j^{(m_1,m_2)}\Lambda^{(m_1,m_2)}=\Lambda^{(m_1,m_2)},
\end{equation}
where
\begin{equation}\label{pism12}
\pi_{ij}\equiv K_{ij}P_{ij}\,,\qquad
\pi_i^{(m_1,m_2)}\equiv K_iP_i^{(m_1,m_2)}\,,
\end{equation}
so that
\begin{equation}\label{KP}
  K_{ij}\Lambda^{(m_1,m_2)}=-P_{ij}\Lambda^{(m_1,m_2)},\qquad
  K_iK_j\Lambda^{(m_1,m_2)}=P_i^{(m_1,m_2)}P_j^{(m_1,m_2)}\Lambda^{(m_1,m_2)}.
\end{equation}
We shall now outline the construction of the projector~$\La^{(m_1,m_2)}$ in terms of the analogous
projectors $\LaBpm$ for the $BC_N$-type spin Calogero model with PSRO~\eqref{l14} with chirality
$\vep=\pm1$ (cf.~\cite{BBB14}). To this end, recall that $\LaBpm$ projects from the Hilbert space
$L^2(\RR^N)\otimes\mc S$ onto spin wavefunctions antisymmetric under particle permutations and
with parity $\pm1$ under $\pi_i^{(m_1,m_2)}$, i.e.,
\begin{equation}\label{m18}
\pi_{ij}\LaBpm=-\LaBpm,\qquad
\pi_i^{(m_1,m_2)}\LaBpm=\pm\LaBpm,
\end{equation}
The projector $\LaBpm$ can then be expressed as
\begin{equation}
\LaBpm
=\frac{1}{2^NN!} \prod\limits_{j=1}^N\Big(1\pm\pi_j^{(m_1,m_2)}\Big)
.\sum\limits_{l=1}^{N!}\varepsilon_l\mc{P}_l \, ,
\label{m19}
\end{equation}
where $\mc{P}_l$ denotes an element of the realization of the permutation group generated by the
operators $\pi_{ij}$ and $\varepsilon_l$ is the signature of $\mc{P}_l$. From Eqs.~\eqref{piLaD}
and~\eqref{m18} we conclude that
\begin{equation}
\Lambda^{(m_1,m_2)}=\Lambda_{\mathrm B,+}^{(m_1,m_2)}+\Lambda_{\mathrm B,-}^{(m_1,m_2)} \,.
\label{m22}
\end{equation}
Indeed, the right-hand side of the latter equation is clearly a projector, since
\[
\Lambda_{\mathrm
  B,+}^{(m_1,m_2)}\Lambda_{\mathrm B,-}^{(m_1,m_2)}=\Lambda_{\mathrm
  B,-}^{(m_1,m_2)}\Lambda_{\mathrm B,+}^{(m_1,m_2)}=0\,,
\]
and it satisfies~\eqref{piLaD} on account of~\eqref{m18}. Thus the space
\[
V\equiv \Lambda^{(m_1,m_2)}(L^2(\RR^N)\otimes\mc S)
\]
decomposes as the direct sum
\begin{equation}
  V=V_{\mathrm B,+} \oplus V_{\mathrm B,-},
  \qquad V_{\mathrm B,\pm}\equiv \Lambda^{(m_1,m_2)}_{\mathrm B,\pm}(L^2(\RR^N)\otimes\mc S).
\label{m23}
\end{equation}
We have already mentioned that, due to the impenetrable nature of the singularities of the
Hamiltonian $H^{(m_1,m_2)}$, its Hilbert space can be taken as the space $L^2(C)\otimes \mc{S}$ of
spin wavefunctions square integrable on the open set $C$ in Eq.~\eqref{l17}. On the other hand,
any point in $\mbb{R}^N$ not lying on the singular subset $x_i\pm x_j=0$, $1\le i<j\le N$, can be
mapped in a unique way to a point in $C$ by a suitable element of the $D_N$ Weyl group, which is
generated by coordinate permutations and sign reversals of an {\em even} number of
coordinates~\cite{Hu90}. Using this fact, it can be shown that $L^2(C)\otimes\mc{S}$ is actually
isomorphic to the space $V$, and $H^{(m_1,m_2)}$ is equivalent to its natural extension to the
latter space which (with a slight abuse of notation) we shall also denote by $H^{(m_1,m_2)}$. With
this identification, in view of Eq.~\eqref{KP} we can write
\begin{equation}\label{HHp}
H^{(m_1,m_2)}=H^{(m_1,m_2)}\Lambda^{(m_1,m_2)}=H'\Lambda^{(m_1,m_2)}\,,
\end{equation}
where $H'$ acts trivially (as the identity) on $\mc S$.

We shall now explain how the spectrum of $H^{(m_1,m_2)}$ can be derived from that of $H'$ using
the previous equation. To this end, note that by Eq.~\eqref{m23} the Hilbert space $V$ is the
closure of the linear subspace spanned by the spin wavefunctions
\begin{equation}
\psi_{\mbf{n},\mbf{s}}^{\ep}(\mbf{x})=\Lambda_{\mathrm B,\ep}^{(m_1,m_2)}
(\phi_{\mbf{n}}(\mbf{x})\ket{\mbf{s}}) \,,\qquad \ep =\pm\,,
\label{m23a}
\end{equation}
where~$\ket{\mathbf s}\equiv\ket{s_1,\dots, s_N}$ is an arbitrary element of the canonical spin
basis. In fact, the wavefunctions~\eqref{m23a} with \emph{fixed} $\ep$ span a subspace whose
closure is the Hilbert space $V_{\mathrm B,\ep}$. Clearly, the functions~\eqref{m23a} are not
linearly independent. Indeed, using \Eq{m18} it is easy to show that these functions satisfy the
relations
\begin{equation}
  \psi_{\mbf{n},\mbf{s}}^{\ep}(\mbf{x}) = 
  - \psi_{\mbf{n}',\mbf{s}'}^{\ep}(\mbf{x}) \, ,\qquad
  \psi_{\mbf{n},\mbf{s}}^{\ep}(\mbf{x})=
  \ep(-1)^{n_i+f(s_i)}\psi_{\mbf{n},\mbf{s}}^{\ep}(\mbf{x}) \, ,
\label{m21}
\end{equation}
where $\mbf{n}'$ and $\mbf{s}'$ are respectively obtained from $\mbf{n}$ and $\mbf{s}$ by
permuting any two of their components (the same for both). Due to these identities, the sets
$\{\psi_{\mbf{n},\mbf{s}}^{+}(\mbf{x})\}$ and $\{\psi_{\mbf{n},\mbf{s}}^{-}(\mbf{x})\}$ are both
linearly independent provided that the following three conditions are imposed on the quantum
numbers $\mbf{n}$ and $\mbf{s}$:
\begin{enumerate}[i)]
\item To avoid overcounting, and for later convenience, we shall order the components of
  $\mathbf n$ as follows:
  \begin{multline*}
    {\bf n}\equiv(\mbf{n}_{\mathrm e}, \mbf {n}_{\mathrm o})=\big(\overbrace{2p_1, \ldots,
      2p_1}^{k_1}, \, \ldots, \, \overbrace{2p_s,
      \ldots, 2p_s}^{k_s},\\
    \overbrace{2q_1+1, \ldots, 2q_1+1}^{l_1}, \, \ldots, \, \overbrace{2q_t+1, \ldots,
      2q_t+1}^{l_t}\big) \, ,
\end{multline*}
where $0\leq s,t \leq N$, $p_1>p_2>\dots>p_s\geqslant0$ and $q_1>q_2>\dots>q_t\geqslant0$.

\item By the second equation in~\eqref{m21}, the allowed values of $s_i$ corresponding to each $n_i$ are
  given by
  \[
  s_i\in
  \begin{cases}
    \{1, 2, \ldots, m_1\} \, , & \text{for even $n_i$}\, , \\[2pt]
    \{ m_1+1, m_1+2, \ldots, m_1+m_2\} \, , & \text{for odd $n_i$}\,,
  \end{cases} 
  \]
  for the set $\{\psi_{\mbf{n},\mbf{s}}^{+}(\mbf{x})\}$, and by
  \[
    s_i\in
      \begin{cases}
        \{ 1, 2, \ldots, m_1\}  \, , &\text{for odd $n_i$}\, , \\[2pt]
        \{m_1+1,m_1 + 2, \ldots, m_1+m_2\}\, , & \text{for even $n_i$}\, . 
      \end{cases}
  \]
  for the set $\{\psi_{\mbf{n},\mbf{s}}^{-}(\mbf{x})\}$.
  
\item If $n_i=n_j$ and $i<j$ we shall take $s_i>s_j$, again to avoid overcounting.
\end{enumerate}
If the above conditions are satisfied, each of the sets
$\{\psi_{\mbf{n},\mbf{s}}^{\ep}(\mbf{x})\}$ ($\ep=\pm$) is a non-orthogonal basis of the
corresponding subspace $V_{\mathrm B,\ep}$, and the union of these sets provides a non-orthogonal
basis of the whole Hilbert space $V$ by Eq.~\eqref{m23}. We shall next show that $H^{(m_1,m_2)}$
leaves invariant each of the subspaces $V_{\mathrm B,\ep}$, and that it acts triangularly on the
corresponding basis $\{\psi_{\mbf{n},\mbf{s}}^{\ep}(\mbf{x})\}$ provided that we (partially) order
it by the total degree~$|\mathbf n|$. Indeed, using Eqs.~\eqref{m22} and~\eqref{HHp}, and taking
into account that $[H',\Lambda_{\mathrm B,\ep}^{(m_1,m_2)}]=0$ we obtain
\begin{equation}
  H^{(m_1,m_2)}\psi_{\mbf{n},\mbf{s}}^{\ep}(\mbf{x})
  =\Lambda_{\mathrm B,\ep}^{(m_1,m_2)}\bigl((H'\phi_{\mbf{n}}(\mbf{x}))\ket{\mbf{s}}\bigr).
\label{m23b}
\end{equation}
From this equation and Eqs.~\eqref{m9} and~\eqref{m23a} it readily follows that
\begin{equation}
H^{(m_1,m_2)}\psi_{\mbf{n},\mbf{s}}^{\ep}(\mbf{x})
=E_{\mbf{n}}'\psi_{\mbf{n},\mbf{s}}^{\ep}(\mbf{x})
+\sum\limits_{\v \mbf{m} \v <\v \mbf{n} \v} 
C_{ \mbf{m} \mbf{n} }\psi_{\mbf{m},\mbf{s}'}^{\ep}(\mbf{x}) \, ,
\label{m24}  
\end{equation}
where the $C_{ \mbf{m} \mbf{n} }$'s are real constants and $\mbf{s}'$ is a permutation of
$\mbf{s}$ such that $(\mathbf m,\mathbf s')$ satisfies conditions i)--iii) above; see~\cite{BFG09}
for more details. By Eq.~\eq{m24}, the action of $H^{(m_1,m_2)}$ on the whole Hilbert space
$V=V_{\mathrm B,+}\oplus V_{\mathrm B,-}$ is the direct sum of two upper triangular actions on
each of the subspaces $V_{\mathrm B,\pm}$. Consequently, the eigenvalues of this operator are given
by
\begin{equation}
E_{\mbf{n},\mbf{s}}^\ep = E_{\mbf{n}}'=a\v \mbf{n} \v + E_0 \, , 
\label{m25}
\end{equation}
where $\ep=\pm$ and $(\mbf{n},\mbf{s})$ satisfies conditions i)--iii) above. Since the RHS of
\Eq{m25} does not depend on $\ep$ and $\mbf{s}$, the eigenvalue associated with the quantum number
$\mbf{n}$ has an {\em intrinsic degeneracy} $d_{\mbf{n}}^{(m_1,m_2)}$ coming from the two possible
chiralities and the spin degrees of freedom. This intrinsic degeneracy is in fact the sum
 \begin{equation}
 d_{\mbf{n}}^{(m_1,m_2)}=d_{\mbf{n},+}^{(m_1,m_2)}
 +d_{\mbf{n},-}^{(m_1,m_2)},
 \label{m26}
\end{equation}
where $d_{\mbf{n},\ep}^{(m_1,m_2)}$ is the number of spin states satisfying conditions i)--iii)
for the given $\mbf{n}$ and $\ep$. Using these conditions we readily obtain~\cite{BBB14}
\begin{equation}
d_{\mbf{n},+}^{(m_1,m_2)}
=\prod\limits_{i=1}^s \binom{m_1}{k_i}\prod\limits_{j=1}^t 
\binom{m_2}{l_j},\qquad d_{\mbf{n},-}^{(m_1,m_2)}=d_{\mbf{n},+}^{(m_2,m_1)},
\label{m27}
\end{equation}
and therefore
\begin{equation}
d_{\mbf{n}}^{(m_1,m_2)}=
\prod\limits_{i=1}^s \binom{m_1}{k_i}\prod\limits_{j=1}^t \binom{m_2}{l_j}+
\prod\limits_{i=1}^s \binom{m_2}{k_i}\prod\limits_{j=1}^t \binom{m_1}{l_j}\,.
\label{m28}
\end{equation}
Thus the spectrum of the $D_N$-type spin Calogero model with PSRO \eq{l15} is given by the RHS of
Eq.~\eq{m25}, where each level possesses an intrinsic degeneracy given by Eq.~\eq{m28}. Of
course, the actual degeneracy of an energy $a|\mathbf n|+E_0$ is the sum
\[
\sum_{|\mathbf n'|= |\mathbf n|}d_{\mathbf n'}^{(m_1,m_2)}\,,
\]
where the sum is over all multiindices $\mathbf n'$ satisfying condition i) above.

It is worth mentioning at this point that the spectrum of the $BC_N$-type spin Calogero model with
PSRO and chirality $\ep$ in Eq.~\eqref{ham} is also given by the RHS of Eq.~\eqref{m25}, with
$E_0$ replaced by~\cite{BBB14}
\[
E_{0,\mathrm B}=E_0+N\be a^2\,.
\]
Moreover, the intrinsic degeneracy of the energy $a|\bn|+E_{0,\mathrm B}$ is given
by~$d_{\bn,\ep}^{(m_1,m_2)}$. It follows from Eq.~\eqref{m26} that the $D_N$ spin Hamiltonian
$H^{(m_1,m_2)}$ is (up to a constant) the direct sum of two $BC_N$-type spin Calogero models of
opposite chiralities with PSRO. Using Eqs.~\eq{m25} and \eq{m28}, the canonical partition function
of the $D_N$-type spin Calogero model with PSRO can be written as
\begin{equation}
  Z^{(m_1,m_2)}(aT)=q^{E_0/a}\sum\limits_{\mbf{n}}
  d_{\mbf{ n}}^{(m_1,m_2)}q^{\v \mbf{n} \v} \, ,\qquad q\equiv\e^{-1/(k_{\mathrm B}T)}\,,
\label{m29}
\end{equation}
where the sum ranges over all multiindices $\mbf{n} $ satisfying condition i) above. Similarly,
the partition functions of the corresponding $BC_N$-type models~\eqref{ham} are given by
\begin{equation}
Z_{B,\pm}^{(m_1,m_2)}(aT)=q^{(E_{0,\mathrm B})/a}\sum\limits_{\mbf{n}}
d_{\mbf{n},\pm}^{(m_1,m_2)}q^{\v \mbf{n} \v}\equiv Z_{B,\mp}^{(m_2,m_1)}\,.
\label{m31}
\end{equation}
From Eq.~\eq{m26}  it then follows that
\begin{align}
  q^{-E_0/a}Z^{(m_1,m_2)}(aT)&=
  q^{-(E_{0,\mathrm B})/a}\big[Z_{B,+}^{(m_1,m_2)}(aT)+Z_{B,-}^{(m_1,m_2)}(aT)\big]\notag\\
  &=q^{-(E_{0,\mathrm B})/a}\big[Z_{B,+}^{(m_1,m_2)}(aT)+Z_{B,+}^{(m_2,m_1)}(aT)\big]\,.
\label{m30}
\end{align}
In order to apply the freezing trick formula~\eqref{m2}, we need only recall the expression for
the partition function $Z$ of the scalar Calogero model of $D_N$-type derived in
Ref.~\cite{BFG09}, namely
\begin{equation}
  q^{-E_0/a}Z(aT)=\big(1+q^N\big)\prod_i(1-q^{2i})^{-1}
  =q^{-(E_{0,\mathrm B})/a}\big(1+q^N\big)Z_{\mathrm B}(aT) \, ,
\label{m13}
\end{equation}
where $Z_{\mathrm B}$ denotes the partition of the scalar Calogero model of $BC_N$ type. Dividing
Eq.~\eqref{m30} by Eq.~\eqref{m13} and applying the analog of the freezing trick
formula~\eqref{m2} for the partition function $\mc{Z}_{\mathrm B,+}^{(m_1,m_2)}$ of the PF spin
chain of $BC_N$ type~\eqref{l14} we finally obtain
\begin{equation}
  \mc{Z}^{(m_1,m_2)}(q)
  =\big(1+q^N\big)^{-1}\left[\mc{Z}_{\mathrm B,+}^{(m_1,m_2)}(q)+\mc{Z}_{\mathrm B,+}^{(m_2,m_1)}(q)
  \right] \,,
\label{m34}
\end{equation}
where from now on we shall use the variable $q=\e^{-1/(k_{\mathrm B}T)}$ in place of $T$. The
partition function $\mc{Z}_{\mathrm B,+}^{(m_1,m_2)}$ can in turn be expressed in terms of the
partition function $\mc{Z}_{\mathrm A,k}^{(m)}(q)$ of the \su{m} PF chain of type~$A$ with $k$
spins with Hamiltonian
\begin{equation}
\mc{H}_A^{(m)}=\sum\limits_{1\leq i<j\leq k}
\frac{1+P_{ij}}{(\rho_i-\rho_j)^2} \, , 
\label{m37}
\end{equation}
where $\rho_i$ is the $i$-th zero of the Hermite polynomial of degree $k$. Indeed, it is shown in
Ref.~\cite{BBB14} that for $m_2>0$ we have
\begin{equation}
  \mc{Z}_{\mathrm B,+}^{(m_1,m_2)}(q)
  =\sum_{k=0}^Nq^{N-k}\qbinom N{k}{q^2} \, \mc{Z}_{\mathrm A,k}^{(m_1)}(q^2) \,
  \mc{Z}_{\mathrm A,N-k}^{(m_2)}(q^2) \qquad(m_2>0), 
\label{m36}
\end{equation}
where the $q$-binomial coefficient $\qbinom N{k}{q^2}$ is defined as 
\begin{equation}
  \qbinom N{k}{q^2} = \frac{(q^2)_N}{(q^2)_k(q^2)_{N-k}}\,,\qquad
  (q^2)_j\equiv\prod_{i=1}^j(1-q^{2i})\,.
\label{m40}
\end{equation}
Combining Eqs.~\eq{m34} and \eq{m36} we finally arrive at the following expression for the
partition function of the $D_N$-type PF chain with PSRO~\eq{l21} in terms of its type~$A$
counterpart as
\begin{equation}
\mc{Z}^{(m_1,m_2)}(q)=
\sum_{k=0}^N f_{N,k}(q) \,   
\mc{Z}_{\mathrm A,k}^{(m_1)}(q^2) \, \mc{Z}_{\mathrm A,N-k}^{(m_2)}(q^2)\qquad(m_2>0), 
\label{m38}
\end{equation}
where $f_{N,k}(q)$ is given by
\begin{equation}
f_{N,k}(q)=\frac{q^{N-k}+q^{k}}{1+q^N}\,\qbinom N{k}{q^2} \,. 
\label{m39}
\end{equation}
The case $m_2=0$, for which $P_i^{(m_1,0)}=1$ and the Hamiltonian~\eqref{l21} reduces to the
rational version of the (trigonometric) Simons--Altshuler chain~\cite{SA94}, deserves special
attention. Indeed, in this case by Eq.~\eqref{m21} the components of the multiindex $\bn$ are all
even (resp.~odd) for the eigenfunctions $\psi_{\mathbf n,\mathbf s}^+$ (resp.~$\psi_{\mathbf
  n,\mathbf s}^-$). As shown in Ref.~\cite{BBB14}, this entails that for $m_2=0$ Eq.~\eqref{m36}
should be replaced by
\begin{equation}\label{cZm10}
\mc{Z}_{\mathrm B,+}^{(m_1,0)}(q)=\mc Z_{\mathrm A,N}^{(m_1)}(q^2)\,.
\end{equation}
On the other hand, since $P_i^{(0,m_2)}=-1$ we have
\[
\mc H^{(0,m_2)}_{\mathrm B,+}=\mc H^{(m_1,0)}_{\mathrm B,+}+\sum_i\frac{2\be}{\xi_i^2}=
\mc H^{(m_1,0)}_{\mathrm B,+}+N
\]
by Eqs.~(A2)-(A5) of Ref.~\cite{BFGR08}. From~\eqref{cZm10} it then follows that
\begin{equation}\label{cZm10final}
\mc{Z}_{\mathrm B,+}^{(0,m_2)}(q)=q^N\mc Z_{\mathrm A,N}^{(m_1)}(q^2)\,,
\end{equation}
and substituting into Eq.~\eqref{m34} we finally obtain
\begin{equation}
  \label{m38b}
  \mc Z^{(m_1,0)}(q)=\mc Z_{\mathrm A,N}^{(m_1)}(q^2)\,.
\end{equation}
Note that, as shown in Ref.~\cite{BBB14}, the RHS of the latter equation also coincides with the
partition function of the $BC_N$-type chain~\eqref{l14} with $\vep=1$ and $m_2=0$. This was to be
expected, as the latter model reduces to its $D_N$ counterpart~\eqref{l21} when $m_2=\be=0$ and
its spectrum does not depend on $\be$.

As is well known, several equivalent closed-form expressions for the partition function of the
$A_{k-1}$-type PF chain \eq{m37} exist in the in the
literature~\cite{Po94,BBHS07,BFGR08epl,BBH10}. For instance, Polychronakos~\cite{Po94} showed that
this function is given by
\begin{equation}
  \label{ZAPoly}
  \mc{Z}_{\mathrm A,k}^{(m)}(q)=\sum_{k_1+\dots +k_m=k}q^{\frac12\sum\limits_{i=1}^mk_i(k_i-1)
  }[k_1,\dots,k_m]_{q}\,,
\end{equation}
where the $q$-multinomial coefficient~$[k_1,\dots,k_m]_{q}$ is defined by
\[
[k_1,\dots,k_m]_{q}=\frac{(q)_{k_1+\dots+k_m}}{\prod\limits_{i=1}^m(q)_{k_i}}\,.
\]
Another well-known expression for the partition function~$\mc{Z}_{\mathrm A,k}^{(m)}$ was derived
in Ref.~\cite{BFGR08epl}, namely
\begin{equation}
\mc{Z}_{\mathrm A,k}^{(m)}(q)
= \sum_{ {\bf f} \in 
\mc{P}_{k}} d_{m}({\bf f})\,q^{\sum\limits_{j=1}^{r-1} \mc{F}_j}
\prod_{j=1}^{k-r}(1- q^{\mc{F}_j'}) \,.
 \label{m41}
\end{equation}
Here $\mc{P}_{k}$ represents the set of all ordered partitions
$\mbf{f} \equiv \{f_1, f_2, \dots, f_r\}$ of the integer~$k$,
$d_{m}({\bf f})= \prod_{i=1}^r \binom{m}{f_i}$, $\mc{F}_j = \sum_{i=1}^j f_i$ are the partial sums
of $\mbf{f}$, and the complementary partial sums are defined as
$\{ \mc{F}_1', \mc{F}_1' , \dots , \mc{F}_{k-r}' \} \equiv \{1, 2, \dots, k\}\setminus\{\mc{F}_1,
\mc{F}_2, \dots, \mc{F}_r\}$.
A related expression for the partition function of the chain \eq{m37} can be obtained by
exploiting its connection with a one-dimensional classical vertex model consisting of $k+1$
vertices connected by $k$ intermediate bonds~\cite{BBH10}. Any possible state for this vertex
model can be represented by a path configuration given by
\begin{equation}
 \vec{s} \equiv \{ s_1,s_2,\dots,s_k\} \, ,
\label{m42}
\end{equation}
where $s_i \in \{1,2,\cdots , m \} $ denotes the spin state of the $i$-th bond. The energy
function associated with this spin path configuration $\vec{s}$ is defined as
\begin{equation}
E^{(m)}(\vec{s}) = \sum_{j=1}^{k-1} j \,
\theta(s_j-s_{j+1}) \, , 
\label{m43}
\end{equation}
where $\theta$ is Heaviside's step function, defined as
\begin{equation}
\theta(x) = \begin{cases}
0\,, & \text{if } x < 0 \, ,
\\[2pt]
1\,, & \text{if }x \geq 0 \,  . 
\end{cases}
\label{m44}
\end{equation}
Using the Yangian quantum group symmetry of the model \eq{m37}, it can be shown that its partition
function coincides with that of the one-dimensional vertex model with energy function~\eq{m43}
(cf.~\cite{BBH10}). Thus $\mc{Z}_{\mathrm A,k}^{(m)}(q)$ can be expressed as
\begin{equation}
\mc{Z}_{\mathrm A,k}^{(m)}(q)=\sum_{\vec{s}} q^{E^{(m)}(\vec{s})} \,, 
\label{m45}
\end{equation}
where the sum has runs over all possible $m^k$ spin path configurations. In particular, from
Eq.~\eqref{m45} it follows that
\[
\mc{Z}_{\mathrm A,k}^{(1)}(q)=q^{\frac12\,k(k-1)}\,.
\]
Thus the partition function~\eqref{m38} with $m_2=1$ reduces to
\begin{equation}
  \mc{Z}^{(m_1,1)}(q)=
  \sum_{k=0}^N q^{(N-k)(N-k-1)}
  f_{N,k}(q) \,   
  \mc{Z}_{\mathrm A,k}^{(m_1)}(q^2) \,.
\label{Zm1}
\end{equation}
It is obvious from any of the expressions~\eqref{ZAPoly}, \eqref{m41} or~\eqref{m45} that the
partition function~$\mc{Z}_{\mathrm A,k}^{(m)}(q)$ is a polynomial in $q$. In particular, from
Eq.~\eqref{m38b} it follows that $\mc Z^{(m_1,0)}$ is an even polynomial in $q$, and its energies
are therefore even nonnegative integers. By Eq.~\eqref{m38}, to show that the partition function
of the $D_N$-type PF chain with PSRO is a polynomial in $q$ when $m_2>0$ it suffices to prove that
the coefficients $f_{N,k}(q)$ in Eq.~\eqref{m39} depend polynomially on $q$. Although it is well
known that the $q$-binomial coefficient $\qbinom N{k}{q^2}$ in \eq{m40} is indeed an even
polynomial in $q$ of degree $2k(N-k)$~\cite{Ci79}, it is not clear whether $f_{N,k}(q)$ is also a
polynomial. In fact, we have verified that this is the case for a wide range of values of $N$ and
all $k\le N$. We conjecture that this is true in general, so that when $m_2>0$ the energies of the
spin chain~\eqref{l21} are also nonnegative integers. Note that the latter fact also follows from
the freezing trick formula~\eqref{m1}, Eq.~\eqref{m25} for the spectrum of the spin dynamical
model~\eqref{l15} and the analogous formula for the scalar $D_N$-type Calogero model in
Ref.~\cite{BFG09}.

\section{Statistical properties of the spectrum}\label{sec.stat}

A characteristic property of {\em all} spin chains of Haldane--Shastry type is the fact that their
level density approaches a Gaussian distribution as the number of spins tends to infinity. This
property has been rigorously proved for the chains of $A_{N-1}$ type and their related
one-dimensional vertex models~\cite{EFG10,BB12}, and has been numerically checked for the $B_N$,
$BC_N$ and $D_N$ type chains with standard spin reversal
operators~\cite{EFGR05,BFGR08,BFG09,BFG13}. More recently, it has been established that the level
density of the $BC_N$-type PF chain with PSRO shows a similar behavior~\cite{BBB14}. It is
therefore of interest to ascertain whether the level density of the $D_N$-type spin chain with
PSRO in Eq.~\eqref{l21} becomes normally distributed as the number of spins tends to infinity. In
fact, Figs.~\ref{fig.31-22} and~\ref{fig.bd} clearly suggest that this is actually the case. We
shall restrict ourselves in the rest of this section to the case $m_2>0$, since for $m_2=0$ the
spectrum of the chain~\eqref{l21} is twice that of an $\mathrm{su}(m_1)$ PF chain of $A_{N-1}$
type (with the same degeneracies) on account of Eq.~\eqref{m38b}.

The spectrum of the spin chain~\eqref{l21} can be determined for any fixed $N$ by evaluating its
partition function~\eqref{m38} with the help of, e.g.,~\textsc{Mathematica}. It turns out that the
most efficient way to compute the partition function $\mc{Z}_{\mathrm A,k}^{(m)}$ appearing in the
latter equation is using the recursion relation
\begin{equation}\label{ZArr}
  \mc{Z}_{\mathrm
    A,k}^{(m)}(q)=\sum_{l=1}^{\min(m,k)}\binom{m}{l}q^{k-l}\prod_{i=1}^{l-1}(1-q^{k-i})\cdot
  \mc{Z}_{\mathrm
    A,k-l}^{(m)}(q)
\end{equation}
with the initial condition $\mc{Z}_{\mathrm A,0}^{(m)}(q)=1$ (see~\ref{app.rr}). In this way it is
possible to evaluate the partition function $\mc{Z}^{(m_1,m_2)}(q)$ on a standard desktop computer
for relatively high values of $N$ (of the order of $50$) and, say, $m_1+m_2\le4$. Our computations
show that the energy levels of the $D_N$-type spin chain with PSRO are always a set of consecutive
integers. This result is consistent with the fact that the spectrum of all previously studied
rational spin chains of HS type is a set of consecutive integers \cite{Po94,BFGR08,BFG09},
including the rational spin chain of $BC_N$ type with PSRO introduced in Ref.~\cite{BBB14}. For
this reason, in order to test the Gaussian character of the level density of the chain~\eqref{l21}
as $N\to\infty$ one can compare directly its normalized level density
\begin{equation}\label{n1}
f(\mc E)=m^{-N}\sum_{i=1}^L d_i\,\de(\mc E-\mc E_i)\,,\qquad m\equiv m_1+m_2\,,
\end{equation}
where $\mc E_1< \cdots < \mc {E}_L$ are the distinct energy levels and $d_i$ is the degeneracy of
$\mc{E}_i$, with the Gaussian distribution
\begin{equation}\label{n2}
g(\mc E)=\frac{1}{\sqrt{2\pi}\si}\,\e^{-\frac{(\mc E-\mu)^2}{2\si^2}}
\end{equation}
with parameters $\mu$ and $\si$ given by the mean and standard deviation of the spectrum,
respectively. More precisely, the level density of the chain~\eqref{l21} is asymptotically Gaussian 
provided that
\[
\frac{d_i}{m^N}\simeq g(\mathcal E_i)\,,\qquad N\gg1.
\]

In order to check the validity of the latter equation for any given $m_1$, $m_2$ and $N$ we need
to compute the corresponding values of $\mu$ and $\si$. We shall next show that, as is the case
with other spin chains of HS type, these parameters can be easily evaluated in closed form
from their definition
\begin{equation} 
\mu= m^{-N} \, \tr\mc{H}^{(m_1,m_2)},\qquad
\sigma^2= m^{-N}\tr\Big[\big(\mc{H}^{(m_1,m_2)}\big)^2\Big]-\mu^2 \,.
\label{n4}
\end{equation}
The traces appearing in \eq{n4} can be computed in essentially the same way as for the $BC_N$-type
PF chain with PSRO~\eq{l14}, using the traces of the spin operators $P_{ij}$, $P_i^{(m_1,m_2)}$
and $\widetilde {P}_{ij}^{(m_1,m_2)}$ given in Ref.~\cite{BBB14}. Proceeding in this way we obtain
\begin{align}
\mu&=\bigg(1+\frac{1}{m}\bigg)\sum_{i\ne j}(h_{ij}+\widetilde{h}_{ij}),
\label{n6}\\
\sigma^2 &=2\left(1-\frac{1}{m^2}\right)\sum_ {i\neq
  j}(h^2_{ij}+{\widetilde h}^2_{ij})+\frac{4}{m^2}(t^2-1)\sum_{ i\neq j}h_{ij}\widetilde{h}_{ij}\,,
\label{n7}
\end{align}
where  $t \equiv m_1-m_2$ and
\[
h_{ij}=(\xi_i-\xi_j)^{-2}\,,\qquad\widetilde{h}_{ij}=(\xi_i+\xi_j)^{-2}\,.
\]
The sums in Eqs.~\eqref{n6}-\eqref{n7} can be evaluated by taking the $\be\to0$ limit of the
corresponding formulas in Appendix~A of Ref.~\cite{BFGR08}. We thus obtain
\begin{align}
\mu&=\frac{1}{2} \left(1+\frac{1}{m}\right)N(N-1),
 \label{n8}\\
\sigma^2&=\frac{1}{36}\left(1-\frac{1}{m^2}\right)N(N-1)(4N+1)+
\frac{1}{4m^2}
N(N-1)(t^2-1) \, .
 \label{n9}
\end{align}
\begin{figure}[h]
  \includegraphics[height=4.2cm]{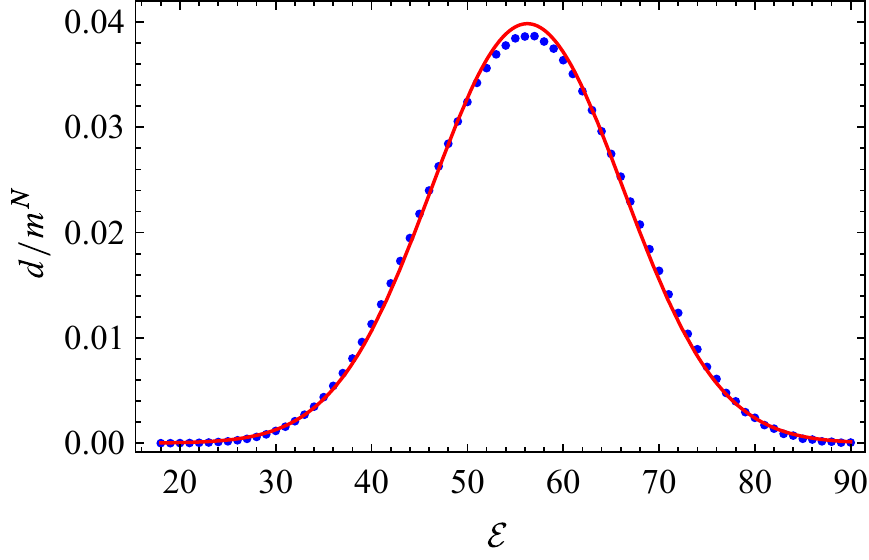}\hfill
  \includegraphics[height=4.2cm]{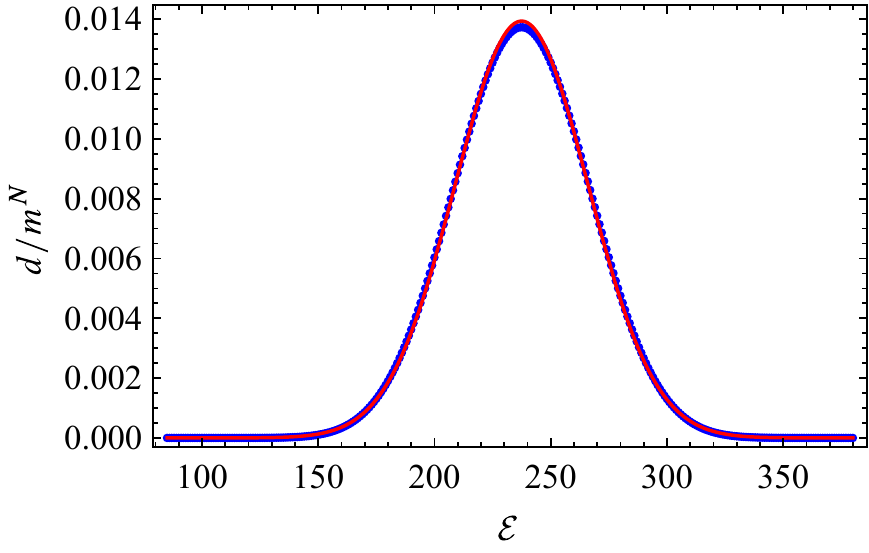}
  \caption{Left: level density of the chain~\eqref{l21} with $m_1=3$, $m_2=1$ and $N=10$ (blue
    dots) compared to the Gaussian distribution~\eqref{n2} (continuous red line). Right: analogous
    plot for $N=20$ spins.}
  \label{fig.ld}
\end{figure}

We have checked that the normalized level density of the spin chain~\eq{l21} is indeed in
excellent agreement with the Gaussian distribution~\eqref{n2} for different values of $m_1$,
$m_2$, and even moderately large values of $N\gtrsim15$. As an example, in Fig.~\ref{fig.ld} we
compare the normalized level density of the chain~\eqref{l21} with $m_1=3$, $m_2=1$, for $N=10$
and $N=20$ spins, respectively, with the corresponding Gaussian distribution~\eqref{n2}. It is
apparent from these plots that the fit, already quite good for $N=10$, improves significantly for
$N=20$. This is confirmed by computing the RMSE errors for both fits, which are respectively equal
to $3.66\times 10^{-2}$ and $2.18\times 10^{-2}$. For comparison purposes, we note that this error decreases
to $1.11\times 10^{-2}$ for $N=50$ spins.

Another interesting property of the spectrum of the chain~\eqref{l21} is connected to the
distribution of the spacings between consecutive levels of the unfolded spectrum~\cite{Ha01},
which in this case is given by
\[
s_i=(\eta_{i+1}-\eta_i)/\De,\qquad i=1,\dots,L-1,
\]
where $\De=(\eta_L-\eta_1)/(L-1)$, $\eta_i=\eta(\mc E_i)$, and
\[
\eta(\mc E)=\int_{-\infty}^{\mc E}g(\mc E')\mathrm d\mc E'=\frac{1}{2}\left[1+\rm{erf}\left(\frac{\mathcal{E}-\mu}
{\sqrt{2}\sigma}\right)\right]\,.
\]
According to a celebrated conjecture due to Berry and Tabor~\cite{BT77}, the distribution of these
spacings for a ``generic'' quantum integrable system should be Poissonian, i.e., $p(s)=\e^{-s}$.
On the other hand, a fundamental conjecture in quantum chaos due to Bohigas, Giannoni and
Schmit~\cite{BGS84} posits that the spacings distribution for a fully chaotic quantum system
invariant under time reversal should follow Wigner's law
\[
p(s)=(\pi s/2)\ms\exp(-\pi s^2/4)\,,
\]
characteristic of the Gaussian orthogonal ensemble in random matrix theory~\cite{Me04}. In fact,
it has been shown that the spacings distribution of a large class of integrable spin chains of
Haldane--Shastry type follows neither Poisson's nor Wigner's law
\cite{FG05,BFGR08,BFGR08epl,BB09,BBB14}. More precisely, it is shown in
Refs.~\cite{BFGR08epl,BFGR08,BFGR09} that the cumulative spacings density
$P(s)\equiv\int_0^sp(s')\mathrm ds'$ of a quantum system with equispaced energy levels and
asymptotically Gaussian level density follows the ``square root of a logarithm law''
\begin{equation}\label{Ps}
  P(s)\simeq 1-\frac{2}{\sqrt{\pi}s_{\mathrm{max}}} \sqrt{\log
    \left(\frac{s_{\mathrm{max}}} {s}\right)}\,,\qquad
  s_{\mathrm{max}}\equiv\frac{\mc E_{L}-\mc E_1}{\sqrt{2\pi}\,\sigma}\,,  
\end{equation}
provided that a few mild technical conditions are satisfied. We have just shown that the energy
levels of the rational $D_N$ chain with PSRO~\eqref{l21} are equispaced and its level density is
asymptotically Gaussian, and it can be easily checked using the formulas for $\mc E_1$ and
$\mc E_L$ below that the technical assumptions in Ref.~\cite{BFGR09} are satisfied. Thus the
spacings distribution of this chain is again approximately given by Eq.~\eqref{Ps}. It should be
noted that for a more precise test of the validity of the Berry--Tabor conjecture one should
restrict oneself to eigenspaces with well-defined quantum numbers corresponding to the main
symmetries of the model. On the other hand, the fact that the spacings distribution of the whole
spectrum is not Poissonian suggests that the Berry--Tabor conjecture does not hold in these
eigenspaces, since the superposition of even a small number of Poissonian distributions is also
Poissonian~\cite{RP60}.

One of the characteristic properties of both the original Haldane--Shastry and the
Polychronakos--Frahm spin chains of $A_{N-1}$ type is their invariance under the quantum group
$Y(\mathrm{sl}(m))$. From the existence of such a large symmetry group one should expect that the
spectrum of these chain exhibits a high degree of degeneracy. In fact, it is shown in
Ref.~\cite{FG15} that the spectrum of these models is far more degenerate than that of a generic
Yangian-invariant system, due to their equivalence to a vertex model of the form~\eqref{m43} with
a very simple dispersion relation. Indeed, as shown in the latter reference, the number
$\nu^{(m)}$ of distinct levels of a generic $Y(\mathrm{sl}(m))$-invariant spin system with a large
number of sites $N$ behaves as $\la_m^N$, where $1<\la_m<2$ is the highest real root of the
polynomial $\la^m-\la^{m-1}-\dots-1$. In contrast, $\nu^{(m)}$ grows as a polynomial in $N$ for
all spin chains of HS type associated with the $A_{N-1}$ root system. For instance, in the case of
the type $A_{N-1}$ PF chain this polynomial is simply given by $\mc E_L-\mc E_1+1$, since its
spectrum is a set of consecutive integers. From the explicit expressions for the maximum and
minimum energies of this model in Ref.~\cite{BFGR08epl} we easily obtain
\begin{equation}\label{nuPFA}
\nu^{(m)}=\frac12\bigg(1-\frac1m\bigg)N^2+\frac{l(m-l)}{2m}+1\qquad (\text{PF chain})\,,
\end{equation}
where $l=N \mod m$. The situation is far less clear for spin chains of HS type associated to other
root systems, with either standard or polarized spin reversal operators. On the one hand, the
presence of these spin reversal operators breaks $\mathrm{su}(m)$ invariance, so that it is not
obvious whether these models are invariant under a suitable quantum group, let alone
$Y(\mathrm{sl}(m))$. On the other hand, it has been observed that the spectrum of some of these
chains is also highly degenerate, which seems to indicate the presence of a large symmetry group.

\begin{figure}[h]
  \centering
  \includegraphics[width=7cm]{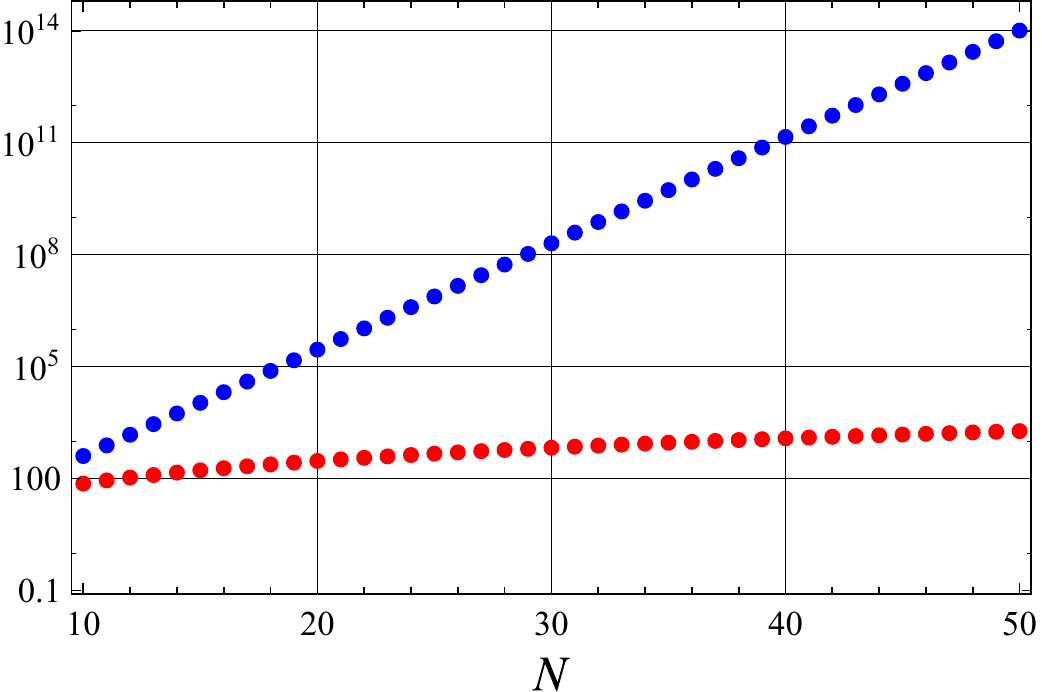}
  \caption{Logarithmic plot of the number of distinct energy levels of the rational $D_N$-type
    chain with PSRO~\eqref{l21} with $m_1=3$, $m_2=1$ (red dots) and of a generic $\mathrm{su}(4)$
    Yangian spin model (red dots) for $10\le N\le 50$.}
\label{fig.degs}
\end{figure}

In the particular case of the $D_N$-type chain with PSRO in Eq.~\eqref{l21}, the number of
distinct energy levels can again be exactly computed under the assumption (which we have
numerically checked) that the spectrum consists of consecutive integers. Indeed, it suffices to
evaluate the maximum and minimum energies $\mc E_{\mathrm{max}}^{(m_1,m_2)}$ and
$\mc E_{\mathrm{min}}^{(m_1,m_2)}$, in terms of which the number $\nu^{(m_1,m_2)}$ of distinct
energy levels is given by
\[
\nu^{(m_1,m_2)}=\mc E_{\mathrm{max}}^{(m_1,m_2)}-\mc E_{\mathrm{min}}^{(m_1,m_2)}+1.
\]
In the first place, the maximum energy can be easily computed by taking into account that $P_{ij}$
and $\tilde P_{ij}^{(m1,m_2)}$ are self-adjoint operators whose square is the identity, so that
their eigenvalues are $\pm1$. Moreover, it is clear that a state of the form $\ket{s,s,\dots,s}$ is
a simultaneous eigenvector of all the operators $P_{ij}$ and $\tilde P_{ij}^{(m1,m_2)}$ with
eigenvalue $1$. Hence the maximum energy of the chain~\eqref{l21} is given by
\begin{equation}\label{cEmaxsum}
\mc{E}_{\mathrm{max}}^{(m_1,m_2)}=
2\sum_{i\neq j}\big[(\xi_i-\xi_j)^{-2}+(\xi_i+\xi_j)^{-2}\big]=N(N-1),
\end{equation}
where the sum was evaluated in~\cite{BFG09}. On the other hand, by Eq.~\eqref{m34} the minimum
  energy is given by
\begin{equation}\label{cEmin2}
 \mc{E}_{\mathrm{min}}^{(m_1,m_2)} 
 =\min \left(\mc{E}^{(m_1,m_2)}_{\mathrm B,+},\mc{E}^{(m_2,m_1)}_{\mathrm B,+}\right),
\end{equation}
where $\mc{E}^{(m_1,m_2)}_{\mathrm B,+}$ is the minimum energy of the $BC_N$-type
chain~\eqref{l14} with $\epsilon=+1$. The latter energy was computed in Ref.~\cite{BBB14}, with
the result
\begin{equation}\label{cEBp}
\mc{E}^{(m_1,m_2)}_{\mathrm B,+}=  (N-l)(N+l-m_1)/m+(l-m_1)\theta(l-m_1)
\end{equation}
where $l\equiv N\mod m$ and $\theta$ is Heaviside's function (cf.~Eq.~\eqref{m44}). Using the above
relation it is straightforward to check that if $m_1\ge m_2$ we have
$\mc{E}^{(m_1,m_2)}_{\mathrm B,+}\le \mc{E}^{(m_2,m_1)}_{\mathrm B,+}$, and therefore
\begin{equation}\label{Emin}
 \mc{E}_{\mathrm{min}}^{(m_1,m_2)}=\mc{E}^{(m_1,m_2)}_{\mathrm B,+},\qquad m_1\ge m_2.
\end{equation}
From Eqs.~\eqref{cEmaxsum} and~\eqref{Emin}, and the assumption that the energy levels are
equispaced, we finally obtain the following closed formula for the number of distinct energy
levels of the $D_N$ chain~\eqref{l21}:
\begin{equation}
  \label{num1m2final}
  \nu^{(m_1,m_2)}=\bigg(1-\frac1m\bigg)N^2-\frac{m_2}m\,N+\frac{l(l-m_1)}m-(l-m_1)\theta(l-m_1)+1\,.
\end{equation}
Thus, it is apparent that $\nu^{(m_1,m_2)}$ is a quadratic polynomial in $N$, as is the case with
the PF chain of $A_{N-1}$ type (cf.~Eq.~\eqref{nuPFA}). In particular, the spectrum of the
chain~\eqref{l21} exhibits a very high degeneracy, much larger than that of a generic
Yangian-invariant $\mathrm{su}(m)$ spin model; see, e.g.,~Fig.~\ref{fig.degs}.

The low number $\nu^{(m_1,m_2)}$ of distinct energy levels of the model~\eqref{l21} entails an
extremely high average degeneracy $d^{(m_1,m_2)}\equiv m^N/\nu^{(m_1,m_2)}$, which in turn
suggests the existence of a large symmetry group. More precisely, it was shown in Ref~\cite{FG15}
that the polynomial growth of the number of distinct energy levels of the spin chains of HS type
associated to the $A_{N-1}$ root system is ultimately due to the equivalence of these chains to a
Yangian-invariant vertex model of the form~\eqref{m43} with a suitable dispersion relation. This
observation makes it reasonable to conjecture that the $D_N$-type spin chain with PSRO~\eqref{l21}
is also invariant under a suitable Yangian group, and that its spectrum coincides with that of a
vertex model analogous to~\eqref{m43} with an appropriate energy function.
\section{The ferromagnetic models}\label{sec.ferro}
We shall consider in this section the ferromagnetic counterparts of the $D_N$-type spin Calogero
model with PSRO~\eq{l15} and its corresponding spin chain~\eq{l21}, with Hamiltonians respectively
given by
\begin{equation}
H_{\mathrm F}^{(m_1,m_2)}=-\sum\limits_{i}
 \frac{\partial^2}{\partial x_i^2}+a\sum\limits_{i\neq j}
\left[\frac{a -P_{ij}}{(x_{ij}^-)^2}+\frac{a-\widetilde
{P}_{ij}^{(m_1,m_2)}}{(x_{ij}^+)^2}\right]
+\frac{a^2}{4}r^2
\label{p2}
\end{equation}
and
\begin{equation} 
\mc{H}_{\mathrm F}^{(m_1,m_2)}=\sum\limits_{i\neq j}
\left[\frac{1- P_{ij}}{(\xi_i-\xi_j)^2}+\frac{1-\widetilde
{P}_{ij}^{(m_1,m_2)}}{(\xi_i+\xi_j)^2}\right] \,.
\label{p1}
\end{equation}
The spectrum of the ferromagnetic spin Calogero model \eq{p2} can be studied in a similar way as
its antiferromagnetic counterpart, following the procedure described in Section~\ref{sec.specPF}. To begin
with, we note that the Hamiltonian \eq{p2} and the auxiliary operator \eq{m3} are related by
\begin{equation}
\label{p3}
H^{(m_1,m_2)}_{\mathrm F}=H'\big|_{K_{ij}
\to P_{ij},\, K_iK_j\rightarrow  P^{(m_1,m_2)}_iP^{(m_1,m_2)}_j}\,.
\end{equation}
Hence, the operator $\Lambda^{(m_1,m_2)}$ in Section~\ref{sec.specPF} should be replaced by the
projector $\Lambda^{(m_1,m_2)}_{\mathrm s}$ onto states {\it symmetric} under simultaneous
exchange of the particles' spatial and spin coordinates, and with parity $+1$ under the product of
an \emph{even} number of operators $\pi_i^{(m_1,m_2)}$ (cf.~\eqref{pism12}). The new projection
operator is the sum
\[
\Lambda^{(m_1,m_2)}_{\mathrm s}=\Lambda_{\mathrm B,\mathrm s,+}^{(m_1,m_2)}+\Lambda_{\mathrm
  B,\mathrm s,-}^{(m_1,m_2)}
\]
of the symmetric analogs of the $BC_N$-type projectors in Section~\ref{sec.specPF}, determined by
\begin{equation}\label{Lasdef}
\pi_{ij}\LaBspm=\LaBspm,\qquad
\pi_i^{(m_1,m_2)}\LaBspm=\pm\LaBspm.
\end{equation}
As explained in Section~\ref{sec.specPF} for the antiferromagnetic case, the operator $H^{(m_1,m_2)}_{\mathrm F}$ is equivalent to
its natural extension to the Hilbert space
\begin{equation}
  V=V_{\mathrm B,\mathrm s,+} \oplus V_{\mathrm B,\mathrm s,-},
  \qquad V_{\mathrm B,\mathrm s,\pm}\equiv \Lambda^{(m_1,m_2)}_{\mathrm B,\mathrm s,\pm}(L^2(\RR^N)\otimes\mc S).
\label{HilF}
\end{equation}
A set of (non-orthogonal) vectors whose linear span is dense in each of the Hilbert spaces
$V_{\mathrm B,\mathrm s,\pm}$ can be constructed in much the same way as in the antiferromagnetic
case, replacing $\LaBpm$ by $\LaBspm$ in~\eqref{m23a}. Due to the symmetry of $\LaBspm$ under
permutations, in order to obtain a basis of these Hilbert spaces we must replace condition iii) in
Section~\ref{sec.specPF} by

iii$'$)\enspace $s_i\ge s_j$ if $n_i=n_j$ and $i<j$\,.

\smallskip\noi As a result, the spectrum of the ferromagnetic model~\eq{p2} is still given by
\Eq{m25}, but the corresponding degeneracy factor $d_{\mbf{n}}^{(m_1,m_2)}$ in \eq{m28} should be
replaced by
\begin{multline}
  d_{\mathrm F,\mbf{n}}^{(m_1,m_2)} =\prod\limits_{i=1}^s\binom{m_1+k_i-1}{k_i}
  \prod\limits_{j=1}^t\binom{m_2+l_j-1}{l_j}\\+ \prod\limits_{i=1}^s\binom{m_2+k_i-1}{k_i}
  \prod\limits_{j=1}^t\binom{m_1+l_j-1}{l_j} \, .
 \label{p5}
\end{multline}
Using this formula for the degeneracy factor and proceeding as in Section~\ref{sec.specPF}, we
find that the partition function of the $D_N$-type ferromagnetic spin chain~\eq{p1} is given by
the following analog of Eq.~\eqref{m34}:
\begin{equation}\label{ZDBp}
\mc{Z}_{\mathrm F}^{(m_1,m_2)}(q)=\big(1+q^N\big)^{-1}\left[\mc{Z}_{\mathrm
    B,\mathrm F,+}^{(m_1,m_2)}(q)+\mc{Z}_{\mathrm B,\mathrm F,+}^{(m_2,m_1)}(q) \right] \,,
\end{equation}
where $\mc{Z}_{\mathrm B,\mathrm F,+}^{(m_1,m_2)}$ denotes the partition function of the
ferromagnetic counterpart of the rational $BC_N$-type chain~\eqref{l14}. Proceeding as in
Ref.~\cite{BBB14} one can readily prove the ferromagnetic version of Eq.~\eqref{m36}, namely
\begin{equation}
 \mc{Z}_{\mathrm B,\mathrm F,+}^{(m_1,m_2)}(q)
  =\sum_{k=0}^Nq^{N-k}\qbinom N{k}{q^2} \, \mc{Z}_{\mathrm A,\mathrm F,k}^{(m_1)}(q^2) \,
  \mc{Z}_{\mathrm A,\mathrm F,N-k}^{(m_2)}(q^2) \qquad(m_2>0)\,.
\label{ZBF}
\end{equation}
Here $\mc{Z}_{\mathrm A,\mathrm F,k}^{(m)}$ denotes the partition function of the ferromagnetic
version of the $\mathrm{su}(m)$ PF~chain of type $A_{N-1}$~\eqref{m37} with $k$ spins, obtained
replacing $P_{ij}$ by $-P_{ij}$ in the latter equation. Finally, from Eqs.~\eqref{ZDBp}
and~\eqref{ZBF} we immediately obtain the following explicit formula for the partition function of
the ferromagnetic chain~\eqref{p1}:
\begin{equation}
  \mc{Z}_{\mathrm F}^{(m_1,m_2)}(q)=
  \sum_{k=0}^N f_{N,k}(q)\, \mc{Z}_{\mathrm A,\mathrm F,k}^{(m_1)}(q^2)\,
  \mc{Z}_{\mathrm A,\mathrm F,N-k}^{(m_2)}(q^2)\qquad(m_2>0)\,,
\label{p6}
\end{equation}
where $f_{N,k}(q)$ is again given by~\eq{m39}. For $m_2=0$, proceeding exactly as in
Section~\ref{sec.specPF} we obtain
\begin{equation}
  \label{ZF0}
  \mc Z^{(m_1,0)}_{\mathrm F}(q)=\mc Z_{\mathrm A,\mathrm F,N}^{(m_1)}(q^2)\,.
\end{equation}

Several explicit expressions for the partition function $\mc Z_{\mathrm A,\mathrm F,k}^{(m)}$ of
the ferromagnetic PF chain of $A_{k-1}$ type appearing in the previous formulas are known in the
literature. The first of these expressions is the analog of Eq.~\eqref{ZAPoly}, namely
\[
\mc{Z}_{\mathrm A,\mathrm F,k}^{(m)}(q)=\sum_{k_1+\dots +k_m=k}[k_1,\dots,k_m]_{q}\,.
\]
Alternatively, $\mc Z_{\mathrm A,\mathrm F,k}^{(m)}$ may be obtained from~\Eq{m41} replacing
$d_{m}({\bf f})$ by its ferromagnetic version
$d_{\mathrm F,m}({\bf f})\equiv \prod_{i=1}^r \binom{m+f_i-1}{f_i}$. Finally,
$\mc Z_{\mathrm A,\mathrm F,k}^{(m)}$ is also given by the RHS of Eq.~\eqref{m45} with $\theta(x)$
replaced by $1-\theta(x)$ in the definition~\eqref{m43} of $E^{(m)}(\vec s)$. From any of these
explicit formulas for $\mc Z_{\mathrm A,\mathrm F,k}(q)^{(m)}$, it follows that this function is a
polynomial in~$q$. By Eqs.~\eqref{p6}-\eqref{ZF0} the same is true for the partition function of
the chain~\eqref{l21}, provided that the coefficient $f_{N,k}(q)$ is a polynomial in $q$.

As is well known, the partition functions of the $A_{N-1}$-type ferromagnetic and
antiferromagnetic PF spin chains satisfy a certain duality relation \cite{Po94,BUW99,HB00}. In
fact, a similar relation also holds for PF chains associated with other root systems
\cite{BFGR08,BFG09,BBB14}. In order to establish a duality relation between the partition
functions of the ferromagnetic and antiferromagnetic spin chains of $D_N$ type with PSRO, it
suffices to observe that their Hamiltonians~\eq{p1} and~\eq{l21} are related by
\begin{equation}
\mc{H}^{(m_1,m_2)}_{\mathrm F}+\mc{H}^{(m_1,m_2)}=2\sum_{i\neq
  j}\big[(\xi_i-\xi_j)^{-2}+(\xi_i+\xi_j)^{-2}\big]=N(N-1)
\label{p9}
\end{equation}
(cf.~\Eq{cEmaxsum}). This obviously implies that the eigenvalues
of~$\mc {H}^{(m_1,m_2)}_{\mathrm F}$ and $\mc {H}^{(m_1,m_2)}$ are also related by~\eq{p9}, so
that their partition functions satisfy the duality relation
\begin{equation}
\mc Z^{(m_1,m_2)}_{\mathrm F}(q)=q^{N(N-1)}\mc Z^{(m_1,m_2)}(q^{-1}).
\label{p10}
\end{equation}

\section{Conclusions and outlook}

We introduce the $D_N$ spin Calogero model with PSRO and its associated spin chain of HS type,
namely the $D_N$ PF chain with PSRO. We solve the former model by finding a suitable
(non-orthonormal) basis of its Hilbert space on which its Hamiltonian acts triangularly. From the
spectrum of this model we are able to compute its partition function in closed form, which yields
the partition function of the spin chain via Polychronakos's freezing trick. More precisely, we
show that the latter partition function can be expressed in terms of the partition function of the
type-$A$ PF chain. Since the type-$A$ partition function can be efficiently evaluated using a
simple recursion formula that we also derive in this paper, we are able to exactly compute the
spectrum of the $D_N$-type chain for relatively high values of $N$. In this way, we are able to
study several global properties of the spectrum of the latter chain. In particular, we provide
strong numerical evidence showing that its energy levels are a sequence of consecutive integers,
and that its level density becomes normally distributed when the number of spins tends to
infinity. From these facts we conclude that the spacings between consecutive levels of the
unfolded spectrum follows a ``square-root-of-a-logarithm'' distribution, characteristic of most
spin chains of HS type. We also determine the number of distinct energy levels of the spin chain,
showing that it is a second-degree polynomial in $N$, as is the case with the PF chain of
$A_{N-1}$ type. For spin chains of HS type related to the $A_{N-1}$ root system, it is
known~\cite{FG15} that the polynomial growth of the number of distinct levels is a consequence of
the fact that these models are equivalent to a Yangian-invariant vertex model with linear energy
function and polynomial dispersion relation. Our results strongly suggest that this is also the
case for the present model, a conjecture which certainly deserves further study. In particular,
the validity of this conjecture would also point out at the existence of a suitable Yangian
symmetry for both the $D_N$-type spin chain and the spin Calogero model with PSRO, as is the case
with the rational and trigonometric Calogero--Sutherland models of $A_{N-1}$-type and their
associated spin chains.

The present work suggests some possible future developments. Among them, the most natural one
would be to address the extension of our results to the Sutherland (both trigonometric and
hyperbolic) models of $BC_N$, $B_N$ and $D_N$ type and their related spin chains. From a more
mathematical standpoint, the fact that the chain's spectrum consists of integers leads us to
conjecture that the function $f_{N,k}(q)$ in Eq.~\eqref{m39} is a polynomial in $q$. Although this
conjecture can be easily checked numerically, we have not been able to find an analytic proof
thereof using the properties of $q$-binomial coefficients.

\section*{Acknowledgments}
This work was partially supported by Spain's MINECO under grant no.~FIS2011-22566, and by the
Universidad Complutense de Madrid and Banco Santander under grant no.~GR3/14-910556.

\appendix

\section{Recursion relation for the partition function of the PF chain of
  $A_{k-1}$~type}\label{app.rr}
We shall provide in this Appendix a short derivation of the recursion relation~\eqref{ZArr}
satisfied by the partition function $\mc Z_{\mathrm A,k}^{(m)}$ of the~$\mathrm{su}(m)$ PF chain
of $A_{k-1}$~type. The main idea behind the proof is to decompose the
multiindex~$\mathbf f\in\mc P_k$ in Eq.~\eqref{m41} as
\[
\mathbf f=(f_1,\dots,f_{r-1},l)\equiv (\tilde{\mathbf f},l)\,,
\]
with $1\le l\le\min(m,k)$ and $\tilde{\mathbf f}\in\mc P_{k-l}$. Setting $s=r-1$ we have
\begin{equation}\label{cFis}
\mc F_1+\cdots+\mc F_{r-1}=\tilde{\mc F}_1+\cdots+\tilde{\mc F}_{s-1}+\mc F_{r-1}=\tilde{\mc
  F}_1+\cdots+\tilde{\mc F}_{s-1}+k-l\,,
\end{equation}
and therefore
\begin{equation}\label{cFisp}
\big\{\mc F'_1,\dots,\mc F'_{k-r}\big\}= \big\{\tilde{\mc F}'_1,\dots\tilde{\mc F}'_{k-l-s}\big\}
\cup\big\{k-l+1,\dots,k-1\big\}\,.
\end{equation}
Substituting~\eqref{cFis} and~\eqref{cFisp} into Eq.~\eqref{m41} we obtain
\begin{align*}
  \mc Z_{\mathrm A,k}^{(m)}(q)&=\sum_{l=1}^{\min(m,k)}\binom
  mlq^{k-l}\prod_{i=1}^{l-1}(1-q^{k-i})\cdot \sum_{\tilde{\mathbf f}\in\mc P_{k-l}}\prod_{i=1}^s
  \binom{m}{\tilde f_i}q^{\tilde{\mc
      F}_1+\cdots+\tilde{\mc F}_{s-1}}\prod_{i=1}^{k-l-s}\big(1-q^{\tilde{\mc F}'_i}\big)\\
  &\equiv \sum_{l=1}^{\min(m,k)}\binom mlq^{k-l}\prod_{i=1}^{l-1}(1-q^{k-i})\cdot \mc Z_{\mathrm
    A,k-l}^{(m)}(q)\,,
\end{align*}
as claimed. As to the initial condition, from Eq.~\eqref{m41} with $k=1$ it easily follows that
$\mc Z_{\mathrm A,1}^{(m)}(q)=m$. From the recursion relation~\eqref{ZArr} with $k=1$ we easily
obtain $\mc Z_{\mathrm A,0}^{(m)}(q)=1$.


\begin{thebibliography}{62}
\expandafter\ifx\csname natexlab\endcsname\relax\def\natexlab#1{#1}\fi
\providecommand{\bibinfo}[2]{#2}
\ifx\xfnm\relax \def\xfnm[#1]{\unskip,\space#1}\fi
\bibitem[{Basu-Mallick et~al.(2014)Basu-Mallick, Bondyopadhaya, and
  Banerjee}]{BBB14}
\bibinfo{author}{B.~Basu-Mallick}, \bibinfo{author}{N.~Bondyopadhaya},
  \bibinfo{author}{P.~Banerjee}, \bibinfo{journal}{Nucl. Phys. B}
  \bibinfo{volume}{883} (\bibinfo{year}{2014}) \bibinfo{pages}{501--528}.
\bibitem[{Polychronakos(1993)}]{Po93}
\bibinfo{author}{A.~P. Polychronakos}, \bibinfo{journal}{Phys. Rev. Lett.}
  \bibinfo{volume}{70} (\bibinfo{year}{1993}) \bibinfo{pages}{2329--2331}.
\bibitem[{Polychronakos(1994)}]{Po94}
\bibinfo{author}{A.~P. Polychronakos}, \bibinfo{journal}{Nucl. Phys. B}
  \bibinfo{volume}{419} (\bibinfo{year}{1994}) \bibinfo{pages}{553--566}.
\bibitem[{Haldane(1988)}]{Ha88}
\bibinfo{author}{F.~D.~M. Haldane}, \bibinfo{journal}{Phys. Rev. Lett.}
  \bibinfo{volume}{60} (\bibinfo{year}{1988}) \bibinfo{pages}{635--638}.
\bibitem[{Shastry(1988)}]{Sh88}
\bibinfo{author}{B.~S. Shastry}, \bibinfo{journal}{Phys. Rev. Lett.}
  \bibinfo{volume}{60} (\bibinfo{year}{1988}) \bibinfo{pages}{639--642}.
\bibitem[{Gutzwiller(1963)}]{Gu63}
\bibinfo{author}{M.~C. Gutzwiller}, \bibinfo{journal}{Phys. Rev. Lett.}
  \bibinfo{volume}{10} (\bibinfo{year}{1963}) \bibinfo{pages}{159--162}.
\bibitem[{Gebhard and Vollhardt(1987)}]{GV87}
\bibinfo{author}{F.~Gebhard}, \bibinfo{author}{D.~Vollhardt},
  \bibinfo{journal}{Phys. Rev. Lett.} \bibinfo{volume}{59}
  (\bibinfo{year}{1987}) \bibinfo{pages}{1472--1475}.
\bibitem[{Gros et~al.(1987)Gros, Joynt, and Rice}]{GJR87}
\bibinfo{author}{C.~Gros}, \bibinfo{author}{R.~Joynt}, \bibinfo{author}{T.~M.
  Rice}, \bibinfo{journal}{Phys. Rev. B} \bibinfo{volume}{36}
  (\bibinfo{year}{1987}) \bibinfo{pages}{381--393}.
\bibitem[{Haldane(1991)}]{Ha91b}
\bibinfo{author}{F.~D.~M. Haldane}, \bibinfo{journal}{Phys. Rev. Lett.}
  \bibinfo{volume}{67} (\bibinfo{year}{1991}) \bibinfo{pages}{937--940}.
\bibitem[{Greiter and Schuricht(2005)}]{GS05}
\bibinfo{author}{M.~Greiter}, \bibinfo{author}{D.~Schuricht},
  \bibinfo{journal}{Phys. Rev. B} \bibinfo{volume}{71} (\bibinfo{year}{2005})
  \bibinfo{pages}{224424(4)}.
\bibitem[{Greiter(2009)}]{Gr09}
\bibinfo{author}{M.~Greiter}, \bibinfo{journal}{Phys. Rev. B}
  \bibinfo{volume}{79} (\bibinfo{year}{2009}) \bibinfo{pages}{064409(5)}.
\bibitem[{Haldane et~al.(1992)Haldane, Ha, Talstra, Bernard, and
  Pasquier}]{HHTBP92}
\bibinfo{author}{F.~D.~M. Haldane}, \bibinfo{author}{Z.~N.~C. Ha},
  \bibinfo{author}{J.~C. Talstra}, \bibinfo{author}{D.~Bernard},
  \bibinfo{author}{V.~Pasquier}, \bibinfo{journal}{Phys. Rev. Lett.}
  \bibinfo{volume}{69} (\bibinfo{year}{1992}) \bibinfo{pages}{2021--2025}.
\bibitem[{Giuliano et~al.(2010)Giuliano, Sindona, Falcone, Plastina, and
  Amico}]{GSFPA10}
\bibinfo{author}{D.~Giuliano}, \bibinfo{author}{A.~Sindona},
  \bibinfo{author}{G.~Falcone}, \bibinfo{author}{F.~Plastina},
  \bibinfo{author}{L.~Amico}, \bibinfo{journal}{New J. Phys.}
  \bibinfo{volume}{12} (\bibinfo{year}{2010}) \bibinfo{pages}{025022(15)}.
\bibitem[{Finkel and Gonz{\'a}lez-L{\'o}pez(2005)}]{FG05}
\bibinfo{author}{F.~Finkel}, \bibinfo{author}{A.~Gonz{\'a}lez-L{\'o}pez},
  \bibinfo{journal}{Phys. Rev. B} \bibinfo{volume}{72} (\bibinfo{year}{2005})
  \bibinfo{pages}{174411(6)}.
\bibitem[{Basu-Mallick and Bondyopadhaya(2006)}]{BB06}
\bibinfo{author}{B.~Basu-Mallick}, \bibinfo{author}{N.~Bondyopadhaya},
  \bibinfo{journal}{Nucl. Phys. B} \bibinfo{volume}{757} (\bibinfo{year}{2006})
  \bibinfo{pages}{280--302}.
\bibitem[{Barba et~al.(2008)Barba, Finkel, Gonz\'alez-L\'opez, and
  Rodr{\'\i}guez}]{BFGR08epl}
\bibinfo{author}{J.~C. Barba}, \bibinfo{author}{F.~Finkel},
  \bibinfo{author}{A.~Gonz\'alez-L\'opez}, \bibinfo{author}{M.~A.
  Rodr{\'\i}guez}, \bibinfo{journal}{Europhys. Lett.} \bibinfo{volume}{83}
  (\bibinfo{year}{2008}) \bibinfo{pages}{27005(6)}.
\bibitem[{Barba et~al.(2009)Barba, Finkel, Gonz\'alez-L\'opez, and
  Rodr{\'\i}guez}]{BFGR09power}
\bibinfo{author}{J.~C. Barba}, \bibinfo{author}{F.~Finkel},
  \bibinfo{author}{A.~Gonz\'alez-L\'opez}, \bibinfo{author}{M.~A.
  Rodr{\'\i}guez}, \bibinfo{journal}{Phys. Rev. E} \bibinfo{volume}{80}
  (\bibinfo{year}{2009}) \bibinfo{pages}{047201(4)}.
\bibitem[{Haldane(1991)}]{Ha91}
\bibinfo{author}{F.~D.~M. Haldane}, \bibinfo{journal}{Phys. Rev. Lett.}
  \bibinfo{volume}{66} (\bibinfo{year}{1991}) \bibinfo{pages}{1529--1532}.
\bibitem[{Kawakami(1992)}]{Ka92}
\bibinfo{author}{N.~Kawakami}, \bibinfo{journal}{Phys. Rev. B}
  \bibinfo{volume}{46} (\bibinfo{year}{1992}) \bibinfo{pages}{1005--1014}.
\bibitem[{Ha and Haldane(1993)}]{HH93}
\bibinfo{author}{Z.~N.~C. Ha}, \bibinfo{author}{F.~D.~M. Haldane},
  \bibinfo{journal}{Phys. Rev. B} \bibinfo{volume}{47} (\bibinfo{year}{1993})
  \bibinfo{pages}{12459--12469}.
\bibitem[{Bernard et~al.(1993)Bernard, Gaudin, Haldane, and Pasquier}]{BGHP93}
\bibinfo{author}{D.~Bernard}, \bibinfo{author}{M.~Gaudin},
  \bibinfo{author}{F.~D.~M. Haldane}, \bibinfo{author}{V.~Pasquier},
  \bibinfo{journal}{J. Phys. A: Math. Gen.} \bibinfo{volume}{26}
  (\bibinfo{year}{1993}) \bibinfo{pages}{5219--5236}.
\bibitem[{Bouwknegt and Schoutens(1996)}]{BS96}
\bibinfo{author}{P.~Bouwknegt}, \bibinfo{author}{K.~Schoutens},
  \bibinfo{journal}{Nucl. Phys. B} \bibinfo{volume}{482} (\bibinfo{year}{1996})
  \bibinfo{pages}{345--372}.
\bibitem[{Kirillov et~al.(1997)Kirillov, Kuniba, and Nakanishi}]{KKN97}
\bibinfo{author}{A.~N. Kirillov}, \bibinfo{author}{A.~Kuniba},
  \bibinfo{author}{T.~Nakanishi}, \bibinfo{journal}{Commun. Math. Phys.}
  \bibinfo{volume}{185} (\bibinfo{year}{1997}) \bibinfo{pages}{441--465}.
\bibitem[{Basu-Mallick et~al.(2008)Basu-Mallick, Bondyopadhaya, and
  Sen}]{BBS08}
\bibinfo{author}{B.~Basu-Mallick}, \bibinfo{author}{N.~Bondyopadhaya},
  \bibinfo{author}{D.~Sen}, \bibinfo{journal}{Nucl. Phys. B}
  \bibinfo{volume}{795} (\bibinfo{year}{2008}) \bibinfo{pages}{596--622}.
\bibitem[{Cirac and Sierra(2010)}]{CS10}
\bibinfo{author}{J.~I. Cirac}, \bibinfo{author}{G.~Sierra},
  \bibinfo{journal}{Phys. Rev. B} \bibinfo{volume}{81} (\bibinfo{year}{2010})
  \bibinfo{pages}{104431(4)}.
\bibitem[{Nielsen et~al.(2011)Nielsen, Cirac, and Sierra}]{NCS11}
\bibinfo{author}{A.~E.~B. Nielsen}, \bibinfo{author}{J.~I. Cirac},
  \bibinfo{author}{G.~Sierra}, \bibinfo{journal}{J. Stat. Mech.-Theory E.}
  (\bibinfo{year}{2011}) \bibinfo{pages}{P11014(39)}.
\bibitem[{Sutherland(1971)}]{Su71}
\bibinfo{author}{B.~Sutherland}, \bibinfo{journal}{Phys. Rev. A}
  \bibinfo{volume}{4} (\bibinfo{year}{1971}) \bibinfo{pages}{2019--2021}.
\bibitem[{Sutherland(1972)}]{Su72}
\bibinfo{author}{B.~Sutherland}, \bibinfo{journal}{Phys. Rev. A}
  \bibinfo{volume}{5} (\bibinfo{year}{1972}) \bibinfo{pages}{1372--1376}.
\bibitem[{Ha and Haldane(1992)}]{HH92}
\bibinfo{author}{Z.~N.~C. Ha}, \bibinfo{author}{F.~D.~M. Haldane},
  \bibinfo{journal}{Phys. Rev. B} \bibinfo{volume}{46} (\bibinfo{year}{1992})
  \bibinfo{pages}{9359--9368}.
\bibitem[{Hikami and Wadati(1993)}]{HW93}
\bibinfo{author}{K.~Hikami}, \bibinfo{author}{M.~Wadati}, \bibinfo{journal}{J.
  Phys. Soc. Jpn.} \bibinfo{volume}{62} (\bibinfo{year}{1993})
  \bibinfo{pages}{469--472}.
\bibitem[{Minahan and Polychronakos(1993)}]{MP93}
\bibinfo{author}{J.~A. Minahan}, \bibinfo{author}{A.~P. Polychronakos},
  \bibinfo{journal}{Phys. Lett. B} \bibinfo{volume}{302} (\bibinfo{year}{1993})
  \bibinfo{pages}{265--270}.
\bibitem[{Calogero(1971)}]{Ca71}
\bibinfo{author}{F.~Calogero}, \bibinfo{journal}{J. Math. Phys.}
  \bibinfo{volume}{12} (\bibinfo{year}{1971}) \bibinfo{pages}{419--436}.
\bibitem[{Frahm(1993)}]{Fr93}
\bibinfo{author}{H.~Frahm}, \bibinfo{journal}{J. Phys. A: Math. Gen.}
  \bibinfo{volume}{26} (\bibinfo{year}{1993}) \bibinfo{pages}{L473--L479}.
\bibitem[{Olshanetsky and Perelomov(1983)}]{OP83}
\bibinfo{author}{M.~A. Olshanetsky}, \bibinfo{author}{A.~M. Perelomov},
  \bibinfo{journal}{Phys. Rep.} \bibinfo{volume}{94} (\bibinfo{year}{1983})
  \bibinfo{pages}{313--404}.
\bibitem[{Bernard et~al.(1995)Bernard, Pasquier, and Serban}]{BPS95}
\bibinfo{author}{D.~Bernard}, \bibinfo{author}{V.~Pasquier},
  \bibinfo{author}{D.~Serban}, \bibinfo{journal}{Europhys. Lett.}
  \bibinfo{volume}{30} (\bibinfo{year}{1995}) \bibinfo{pages}{301--306}.
\bibitem[{Yamamoto and Tsuchiya(1996)}]{YT96}
\bibinfo{author}{T.~Yamamoto}, \bibinfo{author}{O.~Tsuchiya},
  \bibinfo{journal}{J. Phys. A: Math. Gen.} \bibinfo{volume}{29}
  (\bibinfo{year}{1996}) \bibinfo{pages}{3977--3984}.
\bibitem[{Enciso et~al.(2005)Enciso, Finkel, Gonz{\'a}lez-L\'opez, and
  Rodr{{\'\i}}guez}]{EFGR05}
\bibinfo{author}{A.~Enciso}, \bibinfo{author}{F.~Finkel},
  \bibinfo{author}{A.~Gonz{\'a}lez-L\'opez}, \bibinfo{author}{M.~A.
  Rodr{{\'\i}}guez}, \bibinfo{journal}{Nucl. Phys. B} \bibinfo{volume}{707}
  (\bibinfo{year}{2005}) \bibinfo{pages}{553--576}.
\bibitem[{Barba et~al.(2008)Barba, Finkel, Gonz\'alez-L\'opez, and
  Rodr{\'\i}guez}]{BFGR08}
\bibinfo{author}{J.~C. Barba}, \bibinfo{author}{F.~Finkel},
  \bibinfo{author}{A.~Gonz\'alez-L\'opez}, \bibinfo{author}{M.~A.
  Rodr{\'\i}guez}, \bibinfo{journal}{Phys. Rev. B} \bibinfo{volume}{77}
  (\bibinfo{year}{2008}) \bibinfo{pages}{214422(10)}.
\bibitem[{Basu-Mallick et~al.(2009)Basu-Mallick, Finkel, and
  Gonz{\'a}lez-L{\'o}pez}]{BFG09}
\bibinfo{author}{B.~Basu-Mallick}, \bibinfo{author}{F.~Finkel},
  \bibinfo{author}{A.~Gonz{\'a}lez-L{\'o}pez}, \bibinfo{journal}{Nucl. Phys. B}
  \bibinfo{volume}{812} (\bibinfo{year}{2009}) \bibinfo{pages}{402--423}.
\bibitem[{Basu-Mallick et~al.(2011)Basu-Mallick, Finkel, and
  Gonz{\'a}lez-L{\'o}pez}]{BFG11}
\bibinfo{author}{B.~Basu-Mallick}, \bibinfo{author}{F.~Finkel},
  \bibinfo{author}{A.~Gonz{\'a}lez-L{\'o}pez}, \bibinfo{journal}{Nucl. Phys. B}
  \bibinfo{volume}{843} (\bibinfo{year}{2011}) \bibinfo{pages}{505--553}.
\bibitem[{Basu-Mallick et~al.(2013)Basu-Mallick, Finkel, and
  Gonz{\'a}lez-L{\'o}pez}]{BFG13}
\bibinfo{author}{B.~Basu-Mallick}, \bibinfo{author}{F.~Finkel},
  \bibinfo{author}{A.~Gonz{\'a}lez-L{\'o}pez}, \bibinfo{journal}{Nucl. Phys. B}
  \bibinfo{volume}{866} (\bibinfo{year}{2013}) \bibinfo{pages}{391--413}.
\bibitem[{Simons and Altshuler(1994)}]{SA94}
\bibinfo{author}{B.~D. Simons}, \bibinfo{author}{B.~L. Altshuler},
  \bibinfo{journal}{Phys. Rev. B} \bibinfo{volume}{50} (\bibinfo{year}{1994})
  \bibinfo{pages}{1102--1105}.
\bibitem[{Finkel et~al.(2001)Finkel, G\'omez-Ullate, Gonz{\'a}lez-L\'opez,
  Rodr{{\'\i}}guez, and Zhdanov}]{FGGRZ01}
\bibinfo{author}{F.~Finkel}, \bibinfo{author}{D.~G\'omez-Ullate},
  \bibinfo{author}{A.~Gonz{\'a}lez-L\'opez}, \bibinfo{author}{M.~A.
  Rodr{{\'\i}}guez}, \bibinfo{author}{R.~Zhdanov}, \bibinfo{journal}{Commun.
  Math. Phys.} \bibinfo{volume}{221} (\bibinfo{year}{2001})
  \bibinfo{pages}{477--497}.
\bibitem[{Basu-Mallick et~al.(2010)Basu-Mallick, Bondyopadhaya, and
  Hikami}]{BBH10}
\bibinfo{author}{B.~Basu-Mallick}, \bibinfo{author}{N.~Bondyopadhaya},
  \bibinfo{author}{K.~Hikami}, \bibinfo{journal}{SIGMA} \bibinfo{volume}{6}
  (\bibinfo{year}{2010}) \bibinfo{pages}{091(13)}.
\bibitem[{Corrigan and Sasaki(2002)}]{CS02}
\bibinfo{author}{E.~Corrigan}, \bibinfo{author}{R.~Sasaki},
  \bibinfo{journal}{J. Phys. A: Math. Gen.} \bibinfo{volume}{35}
  (\bibinfo{year}{2002}) \bibinfo{pages}{7017--7061}.
\bibitem[{Dunkl(1998)}]{Du98}
\bibinfo{author}{C.~F. Dunkl}, \bibinfo{journal}{Commun. Math. Phys.}
  \bibinfo{volume}{197} (\bibinfo{year}{1998}) \bibinfo{pages}{451--487}.
\bibitem[{Finkel et~al.(2001)Finkel, G\'omez-Ullate, Gonz{\'a}lez-L\'opez,
  Rodr{{\'\i}}guez, and Zhdanov}]{FGGRZ01b}
\bibinfo{author}{F.~Finkel}, \bibinfo{author}{D.~G\'omez-Ullate},
  \bibinfo{author}{A.~Gonz{\'a}lez-L\'opez}, \bibinfo{author}{M.~A.
  Rodr{{\'\i}}guez}, \bibinfo{author}{R.~Zhdanov}, \bibinfo{journal}{Nucl.
  Phys. B} \bibinfo{volume}{613} (\bibinfo{year}{2001})
  \bibinfo{pages}{472--496}.
\bibitem[{Humphreys(1990)}]{Hu90}
\bibinfo{author}{J.~E. Humphreys}, \bibinfo{title}{Reflection Groups and
  Coxeter Groups}, Cambridge Studies in Advanced Mathematics 29,
  \bibinfo{publisher}{Cambridge University Press},
  \bibinfo{address}{Cambridge}, \bibinfo{year}{1990}.
\bibitem[{Basu-Mallick et~al.(2007)Basu-Mallick, Bondyopadhaya, Hikami, and
  Sen}]{BBHS07}
\bibinfo{author}{B.~Basu-Mallick}, \bibinfo{author}{N.~Bondyopadhaya},
  \bibinfo{author}{K.~Hikami}, \bibinfo{author}{D.~Sen},
  \bibinfo{journal}{Nucl. Phys. B} \bibinfo{volume}{782} (\bibinfo{year}{2007})
  \bibinfo{pages}{276--295}.
\bibitem[{Cigler(1979)}]{Ci79}
\bibinfo{author}{J.~Cigler}, \bibinfo{journal}{Monatsh. Math.}
  \bibinfo{volume}{88} (\bibinfo{year}{1979}) \bibinfo{pages}{87--105}.
\bibitem[{Enciso et~al.(2010)Enciso, Finkel, and
  Gonz{\'a}lez-L{\'o}pez}]{EFG10}
\bibinfo{author}{A.~Enciso}, \bibinfo{author}{F.~Finkel},
  \bibinfo{author}{A.~Gonz{\'a}lez-L{\'o}pez}, \bibinfo{journal}{Phys. Rev. E}
  \bibinfo{volume}{82} (\bibinfo{year}{2010}) \bibinfo{pages}{051117(6)}.
\bibitem[{Banerjee and Basu-Mallick(2012)}]{BB12}
\bibinfo{author}{P.~Banerjee}, \bibinfo{author}{B.~Basu-Mallick},
  \bibinfo{journal}{J. Math. Phys.} \bibinfo{volume}{53} (\bibinfo{year}{2012})
  \bibinfo{pages}{083301}.
\bibitem[{Haake(2001)}]{Ha01}
\bibinfo{author}{F.~Haake}, \bibinfo{title}{{Q}uantum {S}ignatures of {C}haos},
  \bibinfo{publisher}{Springer-Verlag}, \bibinfo{address}{Berlin},
  \bibinfo{edition}{second} edition, \bibinfo{year}{2001}.
\bibitem[{Berry and Tabor(1977)}]{BT77}
\bibinfo{author}{M.~V. Berry}, \bibinfo{author}{M.~Tabor},
  \bibinfo{journal}{Proc. R. Soc. London Ser. A} \bibinfo{volume}{356}
  (\bibinfo{year}{1977}) \bibinfo{pages}{375--394}.
\bibitem[{Bohigas et~al.(1984)Bohigas, Giannoni, and Schmit}]{BGS84}
\bibinfo{author}{O.~Bohigas}, \bibinfo{author}{M.~J. Giannoni},
  \bibinfo{author}{C.~Schmit}, \bibinfo{journal}{Phys. Rev. Lett.}
  \bibinfo{volume}{52} (\bibinfo{year}{1984}) \bibinfo{pages}{1--4}.
\bibitem[{Mehta(2004)}]{Me04}
\bibinfo{author}{M.~L. Mehta}, \bibinfo{title}{{R}andom {M}atrices},
  \bibinfo{publisher}{Elsevier}, \bibinfo{address}{San Diego},
  \bibinfo{edition}{3rd} edition, \bibinfo{year}{2004}.
\bibitem[{Basu-Mallick and Bondyopadhaya(2009)}]{BB09}
\bibinfo{author}{B.~Basu-Mallick}, \bibinfo{author}{N.~Bondyopadhaya},
  \bibinfo{journal}{Phys. Lett. A} \bibinfo{volume}{373} (\bibinfo{year}{2009})
  \bibinfo{pages}{2831--2836}.
\bibitem[{Barba et~al.(2009)Barba, Finkel, Gonz\'alez-L\'opez, and
  Rodr{\'\i}guez}]{BFGR09}
\bibinfo{author}{J.~C. Barba}, \bibinfo{author}{F.~Finkel},
  \bibinfo{author}{A.~Gonz\'alez-L\'opez}, \bibinfo{author}{M.~A.
  Rodr{\'\i}guez}, \bibinfo{journal}{Nucl. Phys. B} \bibinfo{volume}{806}
  (\bibinfo{year}{2009}) \bibinfo{pages}{684--714}.
\bibitem[{Rosenzweig and Porter(1960)}]{RP60}
\bibinfo{author}{N.~Rosenzweig}, \bibinfo{author}{C.~E. Porter},
  \bibinfo{journal}{Phys. Rev.} \bibinfo{volume}{120} (\bibinfo{year}{1960})
  \bibinfo{pages}{1698--1714}.
\bibitem[{Finkel and Gonz{\'a}lez-L{\'o}pez(2015)}]{FG15}
\bibinfo{author}{F.~Finkel}, \bibinfo{author}{A.~Gonz{\'a}lez-L{\'o}pez},
  \bibinfo{title}{Yangian-invariant spin models and {F}ibonacci numbers},
  \bibinfo{year}{2015}. \bibinfo{note}{{a}rXiv:1501.05223v1[math-ph]}.
\bibitem[{Basu-Mallick et~al.(1999)Basu-Mallick, Ujino, and Wadati}]{BUW99}
\bibinfo{author}{B.~Basu-Mallick}, \bibinfo{author}{H.~Ujino},
  \bibinfo{author}{M.~Wadati}, \bibinfo{journal}{J. Phys. Soc. Jpn.}
  \bibinfo{volume}{68} (\bibinfo{year}{1999}) \bibinfo{pages}{3219--3226}.
\bibitem[{Hikami and Basu-Mallick(2000)}]{HB00}
\bibinfo{author}{K.~Hikami}, \bibinfo{author}{B.~Basu-Mallick},
  \bibinfo{journal}{Nucl. Phys. B} \bibinfo{volume}{566} (\bibinfo{year}{2000})
  \bibinfo{pages}{511--528}.

\end{thebibliography}

\end{document}